\global\def\draftcontrol{0}

   \def\versionno{ unstable}

\catcode`\@=11

\expandafter\ifx\csname draftcontrol\endcsname\relax\global\def\draftcontrol{0}
\fi

{\count255=\time\divide\count255 by 60
\xdef\hourmin{\number\count255}
\multiply\count255 by-60\advance\count255 by\time
\xdef\hourmin{\hourmin:\ifnum\count255<10 0\fi\the\count255}}
\def\draftdate{\number\month/\number\day/\number\year\ \ \ \hourmin }

\newcommand\makepapertitle{\par
  \begingroup
    \renewcommand\thefootnote{\@fnsymbol\c@footnote}%
    \def\@makefnmark{\rlap{\@textsuperscript{\normalfont\@thefnmark}}}%
    \long\def\@makefntext##1{\parindent 1em\noindent
            \hb@xt@1.8em{%
                \hss\@textsuperscript{\normalfont\@thefnmark}}##1}%
     \newpage
     \global\@topnum\z@   
     \@makepapertitle
     \thispagestyle{empty}\@thanks
  \endgroup
  \setcounter{footnote}{0}%
  \global\let\thanks\relax
  \global\let\makepapertitle\relax
  \global\let\@makepapertitle\relax
  \global\let\@thanks\@empty
  \global\let\@author\@empty
  \global\let\@date\@empty
  \global\let\@title\@empty
  \global\let\title\relax
  \global\let\author\relax
  \global\let\date\relax
  \global\let\and\relax
  \def\version{\let\version\@version\@gobble}
}
\def\@makepapertitle{%
  \newpage
   \ifnum\draftcontrol=1 {}
   \version\versionno
   \vskip 3em%
   \else
   \hfill\hbox to 3cm {\parbox{4cm}{\@pubnum}\hss}%
   \vskip 3em%
   \fi
   \begin{center}%
   \let \footnote \thanks
     {\LARGE {\@title}}%
     \vskip 1.5em%
     {\normalsize
       \lineskip .5em%
       \begin{tabular}[t]{c}%
         \@author
       \end{tabular}\par}%
     \vskip 1.5em%
     {\@bstract}%
     \end{center}%
     \vskip 1.5em
     \@date%
   \par
}

\gdef\@pubnum{}
\def\pubnum#1{%
  \gdef\@pubnum{#1}}

\gdef\@bstract{}
\def\Abstract#1{%
  \gdef\@bstract{%
   \parbox{\textwidth-0pc}{%
   \centerline{\bf Abstract}\penalty1000%
\kern.2cm%
\noindent
\renewcommand\baselinestretch{1.0}%
{#1}}}
}

\def\ps@paper{\let\@mkboth\@gobbletwo%
     \ifnum\draftcontrol=1
    \def\@oddfoot{\hbox to \textwidth{\tiny \versionno \hfil\tiny\draftdate}%
    \hskip -\textwidth \hbox to \textwidth{\hfil\rm\thepage\hfil}}%
     \else\def\@oddfoot{\hbox to \textwidth{\hfil\rm\thepage\hfil}}
     \fi
     \let\@evenfoot\@oddfoot
}

\def\body{\clearpage
          \pagestyle{paper}
    }

\def\@version#1{\ifnum\draftcontrol=1
\typeout{}\typeout{#1}\typeout{}
\vskip3mm\centerline{\hbox{\fbox{\normalsize{\tt DRAFT -- #1 -- }
                   {\draftdate}}}}\vskip3mm
\fi}
\let\version\@version
\long\def\eqlabel#1{\ifnum\draftcontrol=1
                    \tag@false  
                    \tag*{(\theequation) \hbox to -0.2cm{\hspace{0cm}\small{#1}\hss}}
                    \refstepcounter{equation}
                    \edef\@currentlabel{\theequation}
                    \ltx@label{#1}          
                    \else
                    \label{#1}
                    \fi
                    }
\let\st@bibitem\@bibitem
\let\st@lbibitem\@lbibitem
\ifnum\draftcontrol=1
  \def\@bibitem#1{%
    \st@bibitem{#1}\a@@label{#1}\ignorespaces}
  \def\@lbibitem[#1]#2{%
    \st@lbibitem[#1]{#2}\a@@label{#2}\ignorespaces}
  \def\a@@label#1{%
    \gdef\a@lab{\smash{\normalfont\small#1}}
    \ifvmode
      \if@inlabel
        \global\setbox\@labels\hbox{%
          \llap{\a@lab\let\a@lab\relax
                \kern\@totalleftmargin\kern\marginparsep}%
          \box\@labels}%
      \fi
    \fi}
\fi

\documentclass[12pt,letterpaper]{article}

\usepackage{amsmath,amssymb,array,calc,epsfig,rotating,bm}
\usepackage[sort]{cite}
\usepackage{graphicx}
\usepackage{psfrag,verbatim}

\ifnum\draftcontrol=1
\fi

\renewcommand\baselinestretch{1.25}
\renewcommand\section{\@startsection {section}{1}{\z@}%
                                   {-3.5ex \@plus -1ex \@minus -.2ex}%
                                   {2.3ex \@plus.2ex}%
                                   {\normalfont\large\bfseries}}
\renewcommand\subsection{\@startsection{subsection}{2}{\z@}%
                                   {-3.25ex\@plus -1ex \@minus -.2ex}%
                                   {1.5ex \@plus .2ex}%
                                   {\normalfont\normalsize\bfseries}}
\renewcommand\subsubsection{\@startsection{subsubsection}{3}{\z@}%
                                   {-3.25ex\@plus -1ex \@minus -.2ex}%
                                   {1.5ex \@plus .2ex}%
                                   {\normalfont\normalsize\it}}
\renewcommand\paragraph{\@startsection{paragraph}{4}{\z@}%
                                   {-3.25ex\@plus -1ex \@minus -.2ex}%
                                   {1.5ex \@plus .2ex}%
                                   {\normalfont\normalsize\bf}}

\numberwithin{equation}{section}

\def\revise#1       {\raisebox{-0em}{\rule{3pt}{1em}}%
                     \marginpar{\raisebox{.5em}{\vrule width3pt\
                     \vrule width0pt height 0pt depth0.5em
                     \hbox to 0cm{\hspace{0cm}{%
                     \parbox[t]{4em}{\raggedright\footnotesize{#1}}}\hss}}}}

\newcommand\nxt[1]  {\\\fnxt#1}
\newcommand{\ie}{{\it i.e.,}\ }
\newcommand{\eg}{{\it e.g.,}\ }

\def\cala         {{\cal A}}

\def\cale         {{\cal E}}

\def\calh         {{\cal H}}

\def\call         {{\cal L}}
\def\calm         {{\cal M}}
\def\caln         {{\cal N}}
\def\calo         {{\cal O}}

\def\zet          {{\mathbb Z}}

\def\del          {\partial}

\def\Re           {{\rm Re\hskip0.1em}}
\def\Im           {{\rm Im\hskip0.1em}}

\def\sqr#1#2{{\vcenter{\vbox{\hrule height.#2pt
 \hbox{\vrule width.#2pt height#1pt \kern#1pt
 \vrule width.#2pt}\hrule height.#2pt}}}}

\def\a{\alpha}

\def\w{\omega}

\def\c{\chi}

\def\aa1{\phi}
\def\cc1{\psi}

\def\k{\kappa}

\def\l{\lambda}

\def\k{\kappa}

\def\s{\sigma}

\def\ds{d\sigma}

\catcode`\@=12

\begin{document}


\title{\bf Unstable horizons and singularity development in holography}

\date\today

\author{
Pablo Bosch$^{1,4}$,
Alex Buchel$ ^{2,3,4}$ and Luis Lehner$ ^4$\\[0.4cm]
\it $ ^1$Department of Physics \& Astronomy and Guelph-Waterloo Physics
Institute \\University of Waterloo,\\ Waterloo, Ontario N2L 3G1, Canada\\
\it $ ^2$Department of Applied Mathematics,\\
\it $ ^3$Department of Physics and Astronomy\\ 
\it University of Western Ontario\\
\it London, Ontario N6A 5B7, Canada\\
\it $ ^4$Perimeter Institute for Theoretical Physics\\
\it Waterloo, Ontario N2J 2W9, Canada
}

\Abstract{In holographic applications one can encounter
scenarios where a long-wavelength instability can arise.
In such situations, it is often the case that the dynamical 
end point of the instability is a new 
equilibrium phase with a nonlinear scalar hair 
condensate outside the black hole horizon. 
We here review holographic setups where symmetric 
horizons suffer from long-wavelength instabilities where
a suitable equilibrium condensate phase does not exist. 
We study the dynamics of the simplest model in this 
exotic class, and show that it uncovers arbitrarily large
curvatures in the vicinity of the horizon  which asymptotically
turn such region singular, at finite time with
respect to the boundary theory.}

\makepapertitle

\body

\version\versionno
\tableofcontents

\section{Introduction}\label{intro}

The string theory/gauge theory correspondence~\cite{m1,Aharony:1999ti} is by 
now a mature framework 
exploited to address interesting questions in strongly coupled 
gauge theories that are often inaccessible with other theoretical 
tools. In a nutshell, this duality establishes a {\it holographic correspondence}
(a dictionary) between two objects: a non-abelian gauge theory and a higher-dimensional 
gravitational theory/string theory in asymptotically anti de-Sitter space-time. 
One particularly appealing consequence of this correspondence is the fact that 
questions about the 
gauge theory in strongly coupled regimes 
are mapped onto questions in classical gravity. Likewise, a dual gauge theory 
perspective allows for different, and often intuitive, understanding of 
instabilities in black hole/black brane spacetimes.

Indeed, lets recall the physics of {\it holographic superconductors} 
\cite{Hartnoll:2008vx,Hartnoll:2008kx}.
Consider the four-dimensional effective gravitational action\footnote{We set the radius $L$ of 
an asymptotic $AdS_4$ geometry to unity.} in asymptotically 
$AdS_4$ (dual to a three-dimensional conformal field theory $CFT_3$), 
\begin{equation}
S_4=\frac{1}{2\k^2}\int_{\calm_4} dx^4 \left[R+6-\frac 14{F^{\mu\nu}}{F_{\mu\nu}}-\frac 12 
\left(\nabla\phi\right)^2+\phi^2\right] \;.
\eqlabel{s40}
\end{equation}
The four dimensional gravitational constant $\k$ is related to a central charge $c$ of the $CFT_3$ as 
\begin{equation}
c=\frac{192}{\k^2}\; ,
\eqlabel{c}
\end{equation}
$F_{\mu\nu}$ is a field strength of a global $U(1)$ symmetry of the CFT, and $\phi$ is 
a (neutral) gravitational  bulk scalar with 
\begin{equation}
L^2 m^2_\phi=-2 \;,
\eqlabel{mphi}
\end{equation}
which is dual to a dimension $\Delta_\phi=2$ operator $\calo_\phi$ of a boundary theory\footnote{$\phi$
has two alternative quantizations  in $AdS_4$ \cite{Klebanov:1999tb};
our results do not depend on this choice. }. Note that there is  $\zet_2$ symmetry 
in the model, associated with this scalar, $\phi\leftrightarrow -\phi$.
As it is well-known, there are two phases of equilibrium states of this $CFT_3$ at a finite temperature $T$ and a $U(1)$
global symmetry chemical potential $\mu$, distinguished whether $\langle \calo_\phi\rangle =0$ or 
 $\langle \calo_\phi\rangle \ne 0$. The $\langle \calo_\phi\rangle =0$ 
phase exists for arbitrary temperature $T\ge 0$ and it is gravitationally 
described by Reissner-Nordstrom $AdS_4$ black brane with unbroken $\zet_2$ symmetry, correspondingly $\phi\equiv 0$. 
For sufficiently small $T/\mu$ this $\zet_2$ symmetric phase becomes unstable \cite{Hartnoll:2008kx}:
 on the gauge theory side of the correspondence the instability is a generic instability of the order 
parameter in the mean-field theory of thermal second-order phase transitions;
 on the gravity side, this is a Gregory-Laflamme (GL)  type instability \cite{Gregory:1993vy} (in the
sense of being unstable to long-wavelength perturbations)
due to scalarization of the Reissner-Nordstrom $AdS_4$ black brane horizon. 
To understand the gravitational origin of the instability the authors of~\cite{Hartnoll:2008kx} noted
 that even though the scalar 
$\phi$ is above the $AdS_4$ Breitenlohner-Freedman (BF) bound
\begin{equation}
m_\phi^2=-2 \qquad >\qquad m_{BF[AdS_4]}^2=- \frac{(4-1)^2}{4L^2} = -\frac 94\, ,
\eqlabel{abovebf}
\end{equation}
as the Reissner-Nordstrom $AdS_4$ black brane becomes extremal ($T/\mu\to 0$),
it develops $AdS_2\times R^2$ near horizon geometry with the curvature radius
$L_2^2=\frac{L^2}{6}$. In this limit 
\begin{equation}
m_\phi^2=-2 \qquad <\qquad m_{BF[AdS_2]}^2=- \frac{(2-1)^2}{4L_2^2} = -\frac 32\, ,
\eqlabel{belowbf}
\end{equation}
and the bulk scalar $\phi$ becomes unstable (the quasinormal frequency of its linearized fluctuations has $\Im[\w]>0$).
The condensation of the gravitational scalar $\phi$ at low temperatures
is  dynamically saturated by nonlinear effects, spontaneously breaking $\zet_2$ symmetry and 
leading to a new equilibrium phase of the CFT with 
$\langle \calo_\phi \rangle\ne 0$.

There exist many studies and generalizations of the described phenomena in holography\footnote{See \cite{Hartnoll:2009sz}
and references therein.}.
In this paper we focus on a less-known, exotic property of certain black brane/black hole horizons.
As in the example of the holographic superconductor above, imagine a 
holographic\footnote{We point out this feature occurs in top down holographic models, and thus is of importance 
to issues of equilibration and thermalization in strongly coupled gauge theories.} horizon 
with a discrete (or continuous) symmetry. Suppose that there is a critical 
energy\footnote{As we study dynamical phenomena, we work in a microcanonical ensemble.}  
or energy density (for gauge theory states with translational invariance) below which 
the horizon becomes unstable with respect to symmetry breaking (GL) fluctuations.   
There is an equilibrium phase with spontaneously broken symmetry, branching off the GL onset of the instability,
yet, this phase does not exist below the critical energy; moreover, 
it has lower entropy above the criticality than the symmetric phase.
Thus, the horizon representing the symmetric thermal state is unstable, but it is unknown what 
the end point of its instability is. 

To our knowledge, the first realization of the above exotic scenario 
appeared in~\cite{Buchel:2009ge} which was later found  in a top-down 
holographic model in~\cite{Donos:2011ut}. Finally, the same exotic physics is behind the leading instability 
of small black holes in $AdS_5\times S^5$ (dual to $SO(6)$-symmetric states
of strongly coupled $\caln=4$ SYM plasma) 
\cite{Hubeny:2002xn,Buchel:2015gxa,Buchel:2015pla,Dias:2015pda}. Here,
we study the endpoint of this exotic horizon instability.

In the next section we briefly review the bottom-up model of \cite{Buchel:2009ge}.
We discuss the equilibrium states of the system, and the linearized instability 
of symmetric phase states at low energy densities. 
We construct the symmetry-broken phase of the system and demonstrate that it is never
preferred dynamically. In section \ref{dynamic} we employ a characteristic 
formulation of the gravitational dynamics\cite{Chesler:2013lia} in our exotic model. 
We confirm the onset of the GL instability dynamically, and compare the 
linear growth (below the criticality)  and decay (above the criticality) rates 
of the symmetry breaking fluctuations with the corresponding quasinormal mode (QNM) computations
of section \ref{static}.  Next, we present results for the full-nonlinear evolutions of 
unstable horizons. Details of the numerical implementation as well as the convergence 
and the validation of the code are delegated to Appendix \ref{num}.   
We conclude and discuss open questions in section \ref{end}.

\section{Exotic hairy black holes at equilibrium}\label{static}

In this section we review the bottom-up holographic model of the exotic black 
holes presented in~\cite{Buchel:2009ge}. 

The effective four-dimensional gravitational bulk action, dual to a field-theoretic setup 
discussed in the introduction, takes the form
\begin{equation}
\begin{split}
S_4=&S_{CFT}+S_{r}+S_i=\frac{1}{2\kappa^2}\int dx^4\sqrt{-\gamma}\left[\call_{CFT}+\call_{r}+\call_i\right]\,,
\end{split}
\eqlabel{s4}
\end{equation}
\begin{equation}
\call_{CFT}=R+6\,,\qquad \call_r=-\frac 12 \left(\nabla\phi\right)^2+\phi^2\,,\qquad 
\call_i=-\frac 12 \left(\nabla\chi\right)^2-2\chi^2-g \phi^2 \chi^2 \, ;
\eqlabel{lc}
\end{equation}
where we split the action into (a holographic dual to)  a CFT part $S_{CFT}$; its deformation by a relevant
operator $\calo_r$; and a sector $S_i$ involving an irrelevant operator $\calo_i$ 
along with its mixing with $\calo_r$ under the renormalization-group dynamics.
We take bulk quantization so that  the scaling dimension of $\calo_r$ is $\Delta_r=2$;
the scaling dimension of $\calo_i$ is $\Delta_i=4$ . In order to have asymptotically $AdS_4$ solutions, 
we assume that only the normalizable mode of $\calo_i$ is nonzero near the boundary.

The gravitational action \eqref{s4} has $\zet_2\times \zet_2$ discrete symmetry that acts as a parity 
transformation on the scalar fields $\phi$ and $\chi$. The discrete symmetry $\phi\leftrightarrow -\phi$
is explicitly broken by the relevant deformation of the CFT, 
\begin{equation}
\calh_{CFT}\ \to \calh_{CFT}+\Lambda\ \calo_r \;,
\eqlabel{defformation}
\end{equation}
with $\Lambda$ being the deformation mass scale, 
while the $\chi\leftrightarrow -\chi$ 
symmetry is broken spontaneously. The mechanism for the long-wavelength instability 
at play in \eqref{s4} was identified by Gubser \cite{Gubser:2005ih} through the following
observations:
\begin{itemize}
\item consider the linearized dynamics of the $\chi$-sector in the mass-deformed CFT dual to $S_{CFT}+S_r$ 
in \eqref{s4};
\item for the quartic coupling $g<0$, the scalar $\chi$ has an effective mass 
\begin{equation}
m_\c^2=4- 2\ |g|\ \phi^2\,;
\eqlabel{mchi}
\end{equation}
\item homogeneous and isotropic thermal equilibrium states of $S_{CFT}+S_r$ at low temperature (energy densities) 
would result in large values of $\phi$ at the horizon of the dual gravitational description, thus driving 
$m_\c^2$ below the effective BF bound.   
\end{itemize}

A detailed analysis of the homogeneous and equilibrium states of the holographic model \eqref{s4} 
in the canonical ensemble were presented in \cite{Buchel:2009ge}. Here, we present results in the
microcanonical ensemble. We omit all the technical details as the following discussion is a special 
case of the dynamical setup of  section \ref{dynamic}.

\begin{figure}[t]
\begin{center}
\includegraphics[width=2.9in]{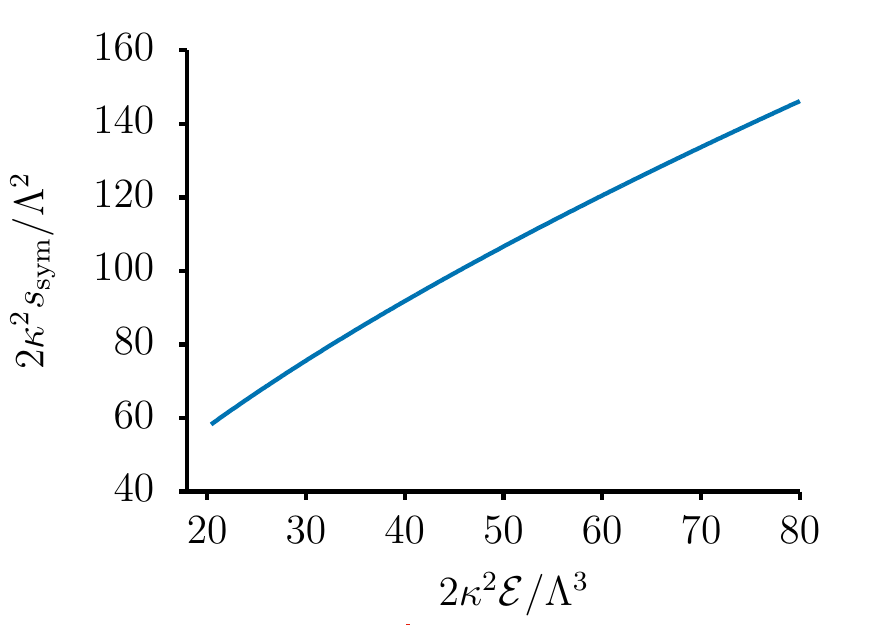} \hfill
\includegraphics[width=2.9in]{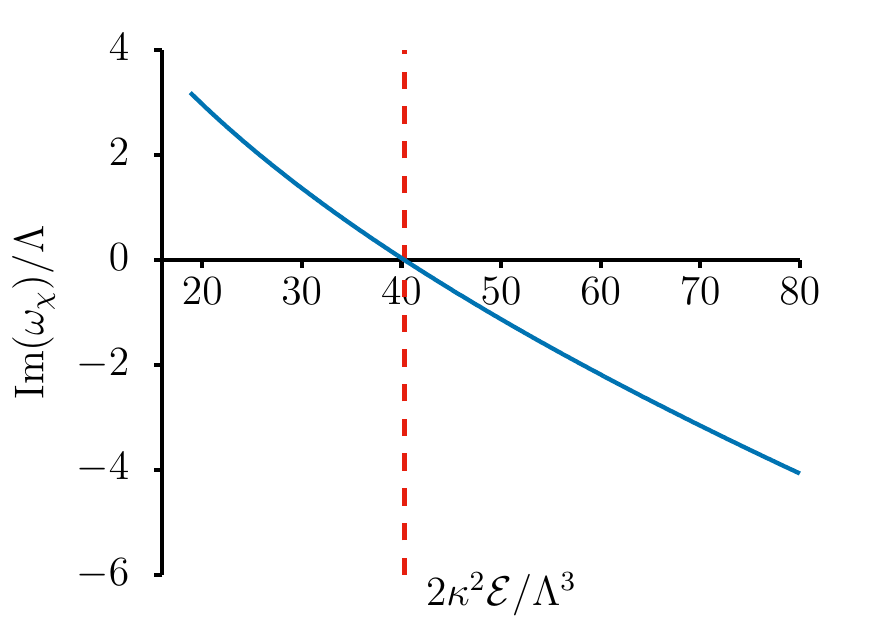}
\end{center}
 \caption{  Entropy density $s_{sym}$ of the $\zet_2$-symmetric phase, \ie with $\langle\calo_i\rangle=0$, 
of exotic black holes  as a function of energy density $\cale$ (left panel).  As the energy density 
is decreased below the critical one $\cale_{crit}$, denoted by a vertical (red)
dashed line, and given that the symmetric phase is perturbatively unstable with
respect to linearized $\zet_2$-symmetry breaking fluctuations,
the imaginary part of the frequency $\w_\chi$ of these fluctuations at zero spatial momenta is 
positive (right panel).} \label{figure1}
\end{figure}

\begin{itemize}
\item There are two equilibrium phases of the holographic model \eqref{s4}, distinguished by the symmetry property 
under $\chi\leftrightarrow -\chi$: the symmetric phase with $\langle\calo_i\rangle=0$, and the symmetry broken phase 
with  $\langle\calo_i\rangle\ne 0$.
\item The entropy density of the symmetric phase $s_{sym}$ as a function of the energy density $\cale$ is presented in 
figure~\ref{figure1}. While this phase is thermodynamically stable 
$\frac{\del^2 \cale}{\del s_{sym}^2}>0$, it is 
perturbatively unstable with respect to a linearized symmetry breaking fluctuations
\cite{Buchel:2010wk}: for $\cale<\cale_{crit}$, with 
\begin{equation}
\frac{2\k^2\cale_{crit}}{\Lambda^3}=40.320(4)\,,
\eqlabel{ecrit}
\end{equation}
the quasinormal modes of the symmetry breaking linearized $\chi$-fluctuations develop a positive imaginary part,
$\Im(\w_{\chi})>0$.  
As emphasized in \cite{Buchel:2010wk}, this model is one of the explicit counterexamples of the Gubser-Mitra 
``correlated stability conjecture'' \cite{Gubser:2000ec,Gubser:2000mm}. 
\end{itemize}

\begin{figure}[t]
\begin{center}
\includegraphics[width=2.9in]{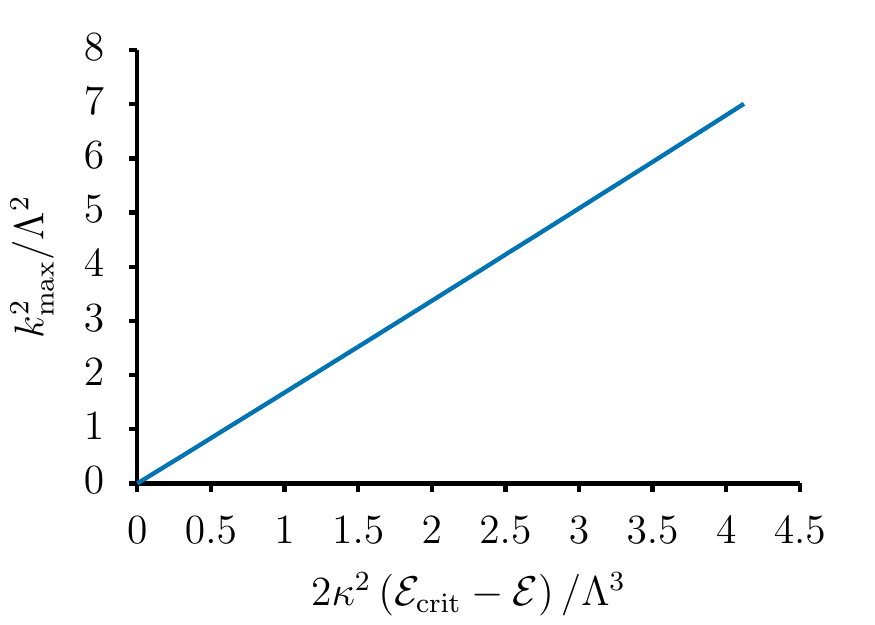} \hfill
\includegraphics[width=2.9in]{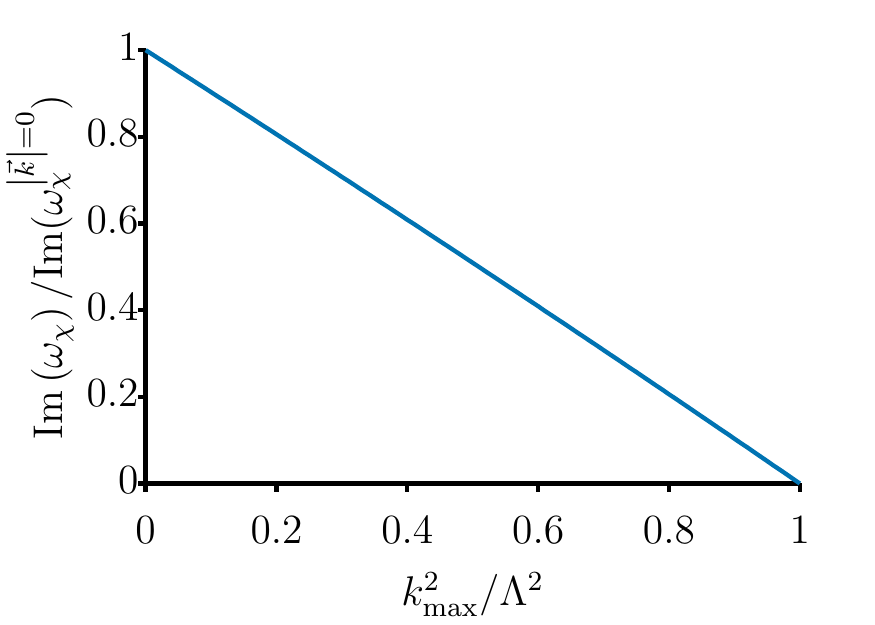}
\end{center}
 \caption{When $\cale<\cale_{crit}$, $\zet_2$-symmetry breaking fluctuations in the symmetric phase of the 
exotic black holes are unstable; the instability persists for the range of the spatial momenta (along the translationally 
invariant directions of the horizon) $\vec{k}$ of the 
fluctuations, $|\vec{k}|\in [0, k_{max}]$. Right panel shows a characteristic dependence
of $\Im(\w_\chi)$ on $|\vec{k}|$ (here ${\cale}/{\cale_{crit}}=0.89780(8)$).
} \label{figure3}
\end{figure}

\begin{itemize}
\item Notice that there is a relation between the behavior of the unstable  $\chi$-mode and
the `GL' instability, in that the $\chi$-instability requires long wavelength modes, 
\ie the instability is cut-off at 
\begin{equation}
|\vec{k}|\le k_{max} \propto (\cale_{crit}-\cale)^{1/2} \;.
\eqlabel{kmax}
\end{equation}  
See figure~\ref{figure3} for further details.
\end{itemize}
However, there is a clear distinction: in the  GL instability~\cite{Gregory:1993vy} the unstable mode is 
hydrodynamic, while the $\chi$-QNM behaves non-hydrodynamically away from the 
critical point, \ie 
$\Im(\w_\chi)\ne 0$ as the spatial momentum (along the translationally invariant directions of the horizon) 
vanishes, $|\vec{k}|=0$.

\begin{figure}[t]
\begin{center}
\includegraphics[width=2.9in]{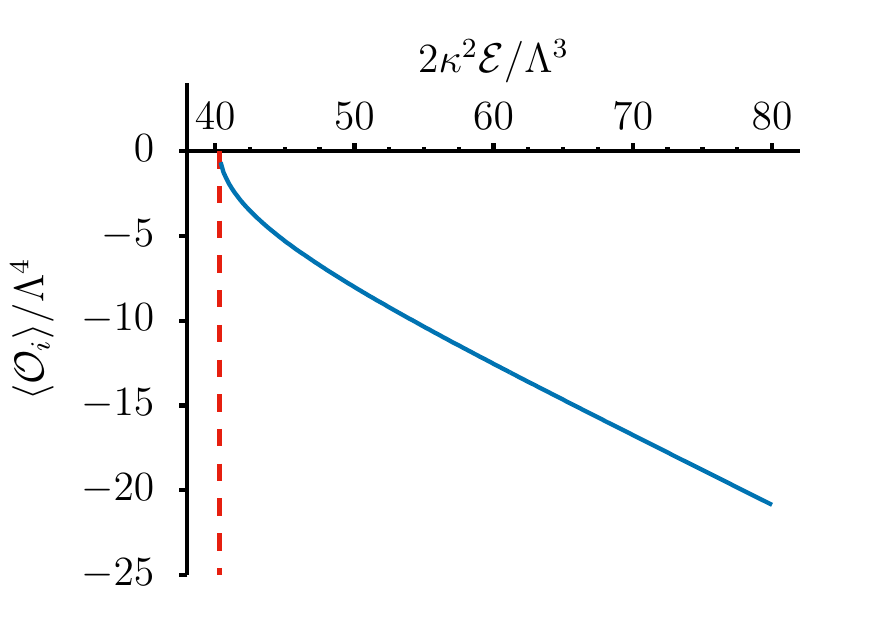} \hfill
\includegraphics[width=2.9in]{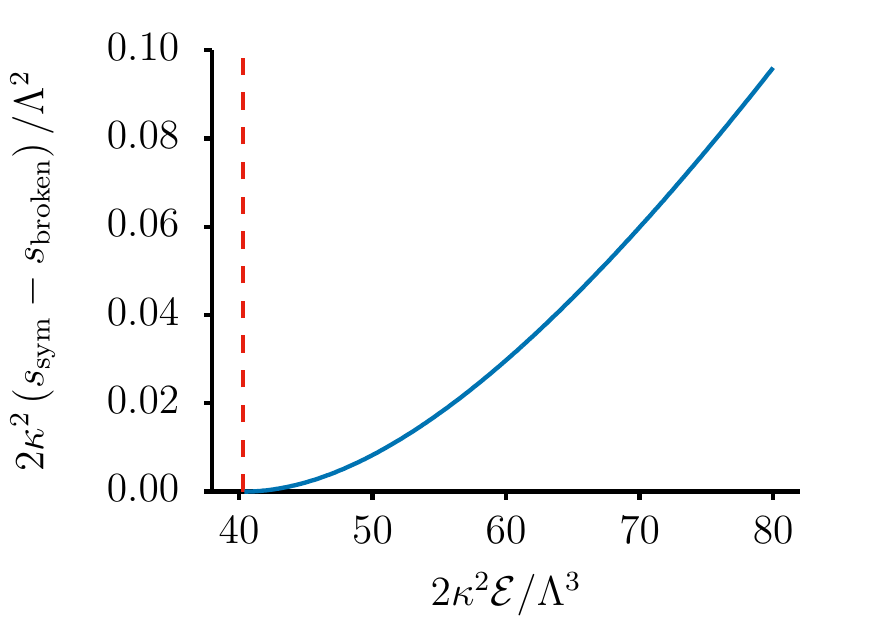}
\end{center}
 \caption{ Exotic black holes have a new equilibrium phase with spontaneously broken $\zet_2$-symmetry 
at energy densities exceeding the critical one, denoted by a vertical (red) dashed line. This phase is characterized 
by $\langle\calo_i\rangle\ne0$, with the expectation value vanishing 
precisely at $\cale=\cale_{crit}$ (left panel). The equilibrium symmetry breaking phase $\langle\calo_i\rangle\ne0$ 
is never realized in a microcanonical ensemble
as its entropy density is always below the corresponding entropy density of the symmetric phase 
(right panel).} \label{figure2}
\end{figure}

\begin{itemize}
\item The expectation value of $\langle\calo_i\rangle\ne 0$ in the symmetry broken phase of the model as a function 
of the equilibrium energy density is presented in figure~\ref{figure2}. This phase exists only for $\cale>\cale_{crit}$,
with 
\begin{equation}
\lim_{\cale\to\cale_{crit}+} \biggl(\langle\calo_i\rangle \ \propto
-\left(\cale-\cale_{crit}\right)^{1/2}\biggr) =0 \;.
\eqlabel{vanoi}
\end{equation}
The equilibrium symmetry broken phase is never realized in a microcanonical
ensemble as it has smaller entropy density compared to the symmetric phase for the same energy density. 
\item Figure~\ref{figure1} exhibits the leading instability at low-energies of the symmetric phase in the 
holographic model \eqref{s4}. In fact, there is a tower of unstable modes (overtones) with critical energies 
$\cale_{crit}^{(n)}$,
\begin{equation}
\cale_{crit}^{(n)} < \cale_{crit}^{(n-1)}\,,\qquad \cale^{(0)}_{crit}\equiv \cale_{crit}\,,\qquad \cale_{crit}^{(1)}\approx 0.26380(6)\ 
\cale_{crit} \;,
\eqlabel{setcrit}
\end{equation}  
parameterized by the number of nodes $(n)$ in the radial profile of the linearized gravitation fluctuations 
$\chi$. Each subleading instability of the symmetric phase identifies a branch point of a new unstable phase
with $\langle\calo_i\rangle\ne 0$. Properties of these new phases are analogous to the broken 
phase in figure~\ref{figure2}, see also 
\cite{Buchel:2009ge}.
\end{itemize}

\begin{figure}[t]
\begin{center}
\includegraphics[width=2.9in]{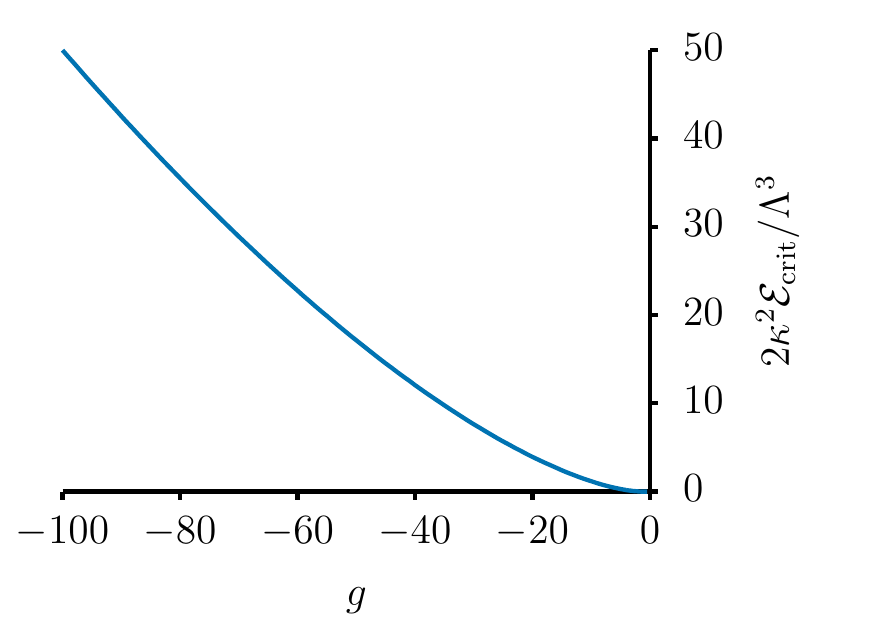} \hfill
\includegraphics[width=2.9in]{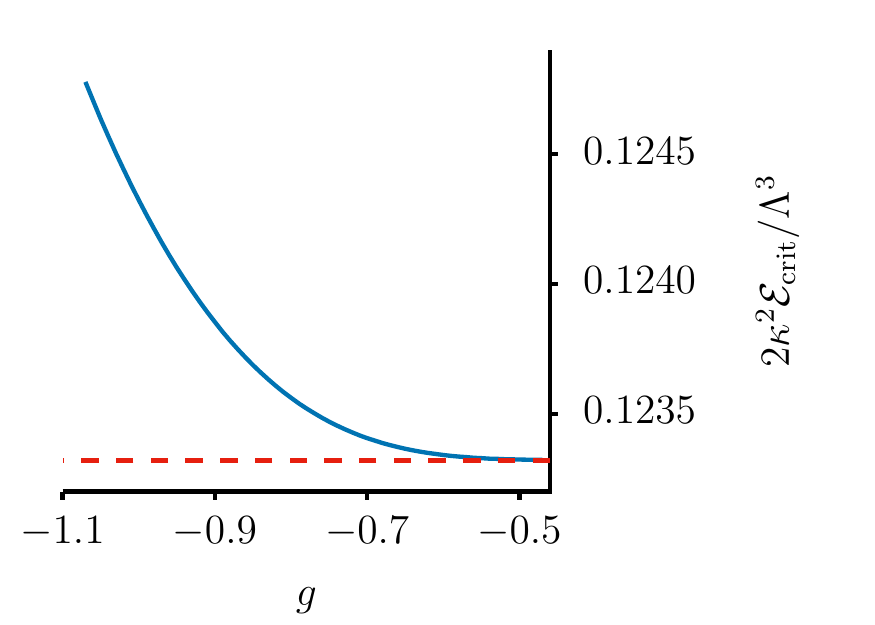}
\end{center}
 \caption{Critical energy density of the leading instability of the symmetric phase as a function of 
the nonlinear coupling $g$. It appears that the instability persists in the limit $g\to 0_-$ (right panel). The (red) 
dashed line identifies the vacuum energy of the symmetric phase, see \eqref{evac}.} \label{figure4}
\end{figure}

\begin{figure}[t]
\begin{center}
\includegraphics[width=2.9in]{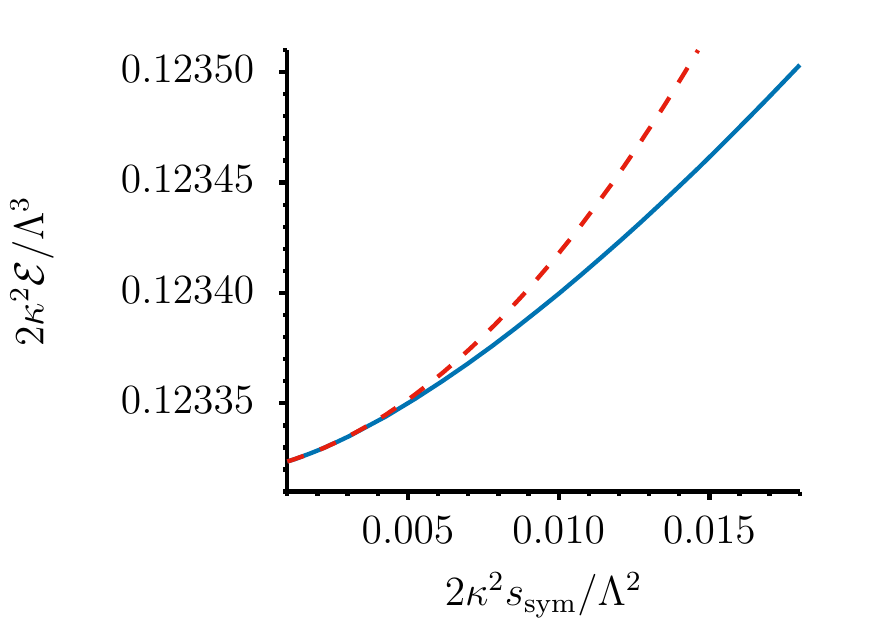} \hfill
\includegraphics[width=2.9in]{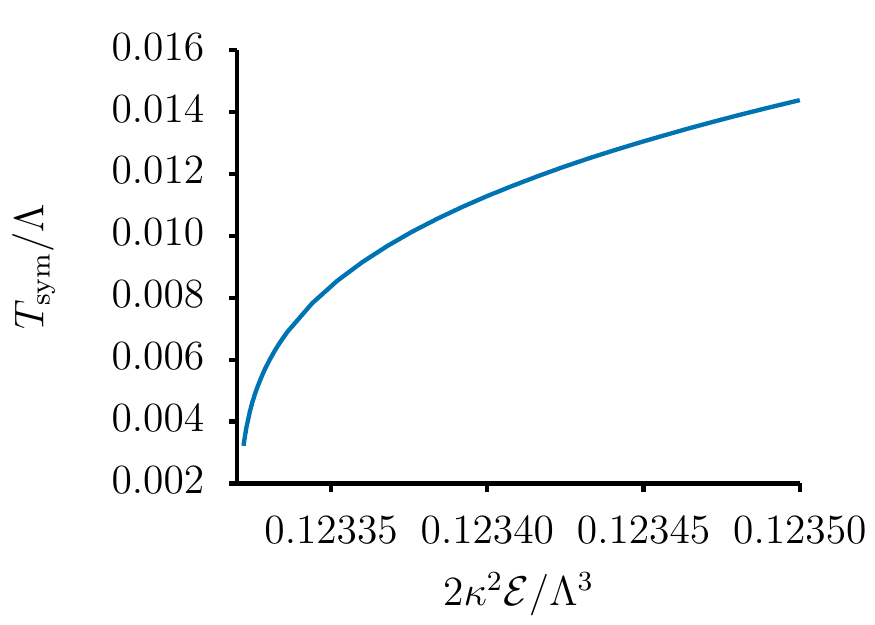}
\end{center}
 \caption{The left panel shows the energy density of the symmetric phase as a function of 
the entropy density. The (red) dashed line is the extrapolation of the energy-entropy data (solid blue line)
in the limit $s_{sym}\to 0$. The extrapolation is used to estimate the vacuum energy of the symmetric phase 
\eqref{evac}. The right panel shows the dependence of the symmetric phase black hole temperature as a function of the 
energy density.} \label{figure5}
\end{figure}

\begin{itemize}
\item The analysis reported above was performed with the nonlinear coupling in the effective action 
\eqref{s4} set to $g=-100$. The phase diagram of the model does not change as $g$ changes, as long 
as $g<0$, see figure~\ref{figure4}. The (red) dashed line in the right panel represents the estimate 
for the vacuum energy of the symmetric phase:
\begin{equation}
\frac{2\k^2\cale^{vacuum}}{\Lambda^3}=0.1233(2) \;.
\eqlabel{evac} 
\end{equation}
Notice that $\cale_{sym}^{vacuum}\to 0$ in the conformal limit $\Lambda\to 0$; to obtain the better 
estimate for  $\cale^{vacuum}$ we extended the analysis of the symmetric phase to the low-entropy 
region, as shown in figure~\ref{figure5}, and extrapolated the energy-entropy data to zero entropy density
(indicated by the (red) dashed line).  The right panel shows the dependence of the temperature $T_{sym}$ of the symmetric 
phase black hole --- the limit $\cale\to \cale^{vacuum}$ appears to correspond to an extremal limit. 
\end{itemize}

In this section we focused on the static phase diagram, along with the linearized (in)stabilities of these phases,
 of the holographic action \eqref{s4},  dual to non-conformal  $QFT_3$ in
 Minkowski space-time $R^{1,2}$. In what follows we will discuss the dynamical case. 
In section \ref{end},  we comment on properties of the model with $QFT_3$ residing in $R\times S^2$. 
Additionally, we comment on the extension of the model \eqref{s4} with the gravitational potential for the   
scalar $\chi$  bounded from below.

\section{Dynamics of the exotic unstable horizons}\label{dynamic}

In this section we discuss dynamical properties of the holographic model \eqref{s4}, with the 
boundary $QFT_3$ formulated in $R^{1,2}$. We follow closely  the holographic numerical framework
in the characteristic formulation as described in, \eg\cite{Chesler:2013lia}.

\subsection{Dynamical setup}

We assume translational invariance along the spatial directions of the boundary. The relevant fields
are described by 
\begin{equation}
\begin{split}
&ds_4^2= 2 dt \left(dr -A(t,r)\ dt \right)+\Sigma(t,r)^2\
\left[dx_1^2+dx_2^2\right] \;, \\
&\phi=\phi(t,r)\,,\qquad \chi=\chi(t,r) \;.
\end{split}
\eqlabel{bgr}
\end{equation}
Einstein equations define the following evolution equations of motion:
\begin{equation}
\begin{split}
&0=d_+'\Sigma+d_+\Sigma\ \left(\ln\Sigma\right)'-\frac 32 \Sigma-\frac 14\Sigma\left(\phi^2-2\chi^2
-g\phi^2\chi^2\right) \;,\\
&0=d_+'\phi+d_+\phi\ \left(\ln\Sigma\right)'+\frac{d_+\Sigma}{\Sigma}\ \phi'
+\phi\left(1-g \chi^2\right)\;,\\
&0=d_+'\chi+d_+\chi\ \left(\ln\Sigma\right)'+\frac{d_+\Sigma}{\Sigma}\ \chi'
-\chi\left(2+g \phi^2\right)\;,\\
&0=A''-2\frac{d_+\Sigma}{\Sigma^2}\ \Sigma'+\frac 12 d_+\phi\ \phi'+\frac 12
d_+\chi\ \chi' \;,
\end{split}
\eqlabel{evolveoms}
\end{equation}
together with the constraint equations:
\begin{eqnarray}
0 &=&\Sigma''+\frac 14\Sigma\left( (\phi')^2+(\chi')^2\right) \label{coneoms1}
\;,\\
0 &=&d_+^2\Sigma-2 A d_+'\Sigma-\frac{d_+\Sigma}{\Sigma^2}\
\left(A\Sigma^2\right)'\nonumber \\
& & +\frac 14\Sigma \left( (d_+\phi)^2+(d_+\chi)^2 
 +2A\left(6+\phi^2-2\chi^2-g\phi^2\chi^2 \right) \right)\;, \label{coneoms2}
\end{eqnarray}
where $'\equiv\del_r$ and $d_+\equiv \del_t+A\ \del_r$.
The constraint equations are preserved by the evolution equations provided they
are satisfied at a given timelike 
surface (\eg~\cite{1962RSPSA.269...21B,Bishop:1997ik,2012LRR....15....2W}) 
--- which in our case is the AdS boundary.  

The general asymptotic boundary  ($r\to\infty$) solution  of the equations of motion, given by
\begin{eqnarray} \label{basym}
\Sigma&=&r+\l(t)-\frac 18 p_1^2\ \frac 1r +\calo\left(\frac{1}{r^2}\right)
\nonumber \;,\\
A&=&\frac{r^2}{2}+\l(t)\ r -\frac 18 p_1^2+\frac 12\l(t)^2
-\dot\l(t) \nonumber \\
& & + \left(\mu-\frac 14 p_1 p_2(t)-\frac 14 p_1^2\l(t)\right)\frac 1r
+\calo\left(\frac{1}{r^2}\right) \nonumber \;, \\
\phi&=&\frac{p_1}{r}+\frac{p_2(t)}{r^2}+\calo\left(\frac{1}{r^3}\right)\, , \nonumber \\
\chi&=&\frac{q_4(t)}{r^4}+\calo\left(\frac{1}{r^5}\right) \, ,
\end{eqnarray}
is characterized by two constants $\{p_1,\mu\}$, and three dynamical variables 
$\{p_2(t), q_4(t), \l(t)\}$. These parameters have the following interpretation:
\begin{itemize}
\item $p_1$ and $p_2(t)$ are correspondingly the non-normalizable and normalizable coefficients
of the bulk scalar $\phi$, identified with the deformation mass scale $\Lambda$ and the 
expectation value of the relevant operator $\calo_r$ of the dual $QFT_3$,
\begin{equation}
p_1=\Lambda\,,\qquad
p_2(t)=\langle\calo_r(t)\rangle\,;
\eqlabel{phidat}
\end{equation}
\item $q_4(t)$ is the normalizable coefficient of the bulk scalar $\chi$, identified 
with the expectation value of the $\zet_2$-symmetry breaking irrelevant operator $\calo_i$
of the dual $QFT_3$,
\begin{equation}
q_4(t)=\langle\calo_i(t)\rangle\,; 
\eqlabel{chidat}
\end{equation}
\item $\mu$ is related to the conserved energy density $\cale$ of the boundary $QFT_3$ as follows
\begin{equation}
\frac{2\k^2 \cale}{\Lambda^3}=\frac{-4\mu}{\Lambda^3} \;;
\eqlabel{endata}
\end{equation}
\item   $\l(t)$ is the residual radial coordinate diffeomorphisms parameter 
\begin{equation}
r\to r+\l(t) \;,
\eqlabel{resdiffeo}
\end{equation}
which can adjusted to keep the apparent horizon at a fixed location, which in our
case will be $r=1$:
\begin{equation}
\biggl(\del_t +A(t,r)\ \del_r\ \biggr) \Sigma(t,r)\ \equiv\
 d_+\Sigma(t,r)\bigg|_{r=1}=0 \;.
\eqlabel{ldata}
\end{equation}
\end{itemize}

To initialize evolution at $t=0$, we  provide the bulk scalar profiles,
\begin{equation}
\phi(t=0,r)=\frac{p_1}{r}+\calo\left(\frac{1}{r^2}\right)\,,\qquad
\chi(t=0,r)=\calo\left(\frac{1}{r^4}\right) \;,
\eqlabel{initphichi}
\end{equation} 
along with the values of $\{p_1,\mu\}$, specifying the dual $QFT_3$ mass scale $\Lambda$ \eqref{phidat} and the initial state energy density  
$\cale$ \eqref{endata}.
The constraint equation \eqref{coneoms1} is then used to determine an initial profile $\Sigma(t=0,r)$. 
Eqs.~\eqref{evolveoms} are then employed to evolve such data \eqref{initphichi} in time. 
The second constraint \eqref{coneoms2}, representing the conservation of the energy density, is enforced requiring that 
a parameter $\mu$ in the asymptotic expansion of $A$, see \eqref{basym}, is time-independent. 

Details of the numerical implementation, specific choices of the initial conditions \eqref{initphichi} used, 
and code convergence tests can be found in Appendix \ref{num}.

\subsection{Dynamics of the symmetric sector}\label{dynsym}

To study dynamics in the {\em symmetric sector}, we adopt initial conditions 
as described in Appendix \ref{initial} with $\cala_p\ne 0$ and $\cala_q=0$,
implying that (in $\l_0\equiv 0$ gauge) 
\begin{equation}
\phi\bigg|_{t=0}=\frac{p_1}{r}+\cala_p\ \frac{\exp\left(-\frac
1r\right)}{r^2}\,,\qquad \chi(t,x)\equiv 0 \;.
\eqlabel{pqinitial}
\end{equation}

\begin{figure}[t]
\begin{center}
\includegraphics[width=2.9in]{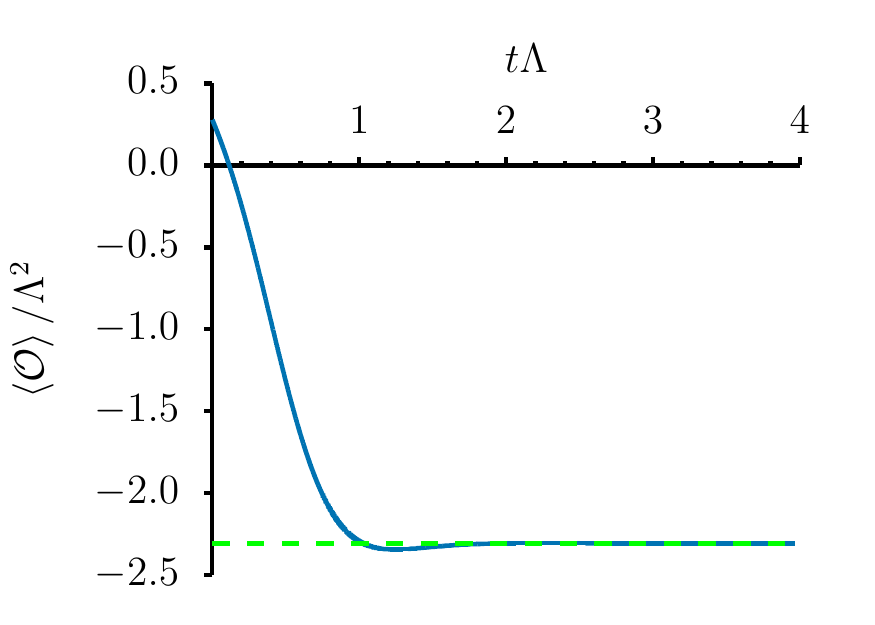} \hfill
\includegraphics[width=2.9in]{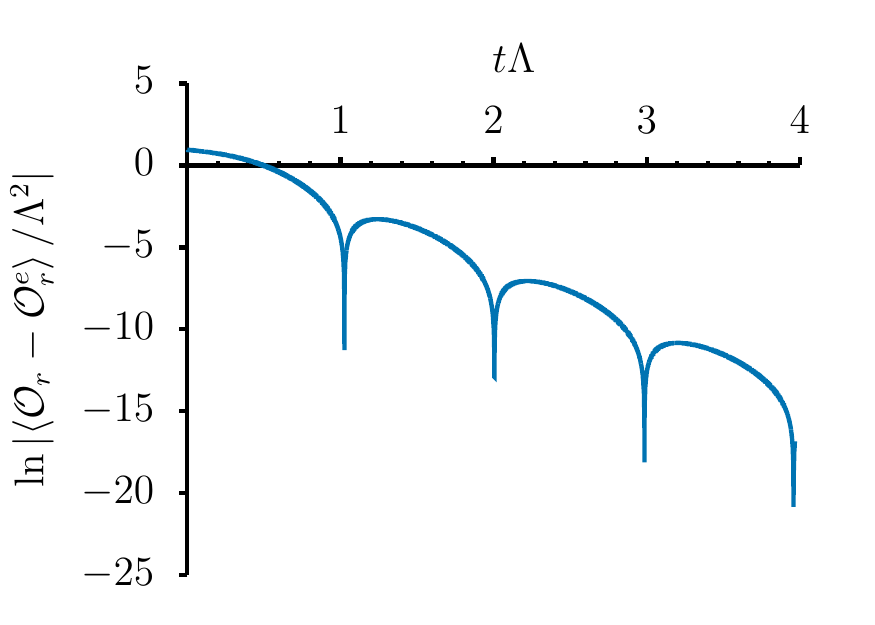}
\end{center}
 \caption{Time evolution of a typical state in $\zet_2$-symmetric phase of  exotic black holes. 
The (green) dashed line is the equilibrium value $\langle \calo_r^e\rangle$ of the operator 
$\calo_r$. Right panel shows that approach towards equilibrium occurs in 
characteristic quasinormal mode ringing of the black hole horizon. } \label{figure6}
\end{figure}

\begin{figure}[t]
\begin{center}
\includegraphics[width=2.9in]{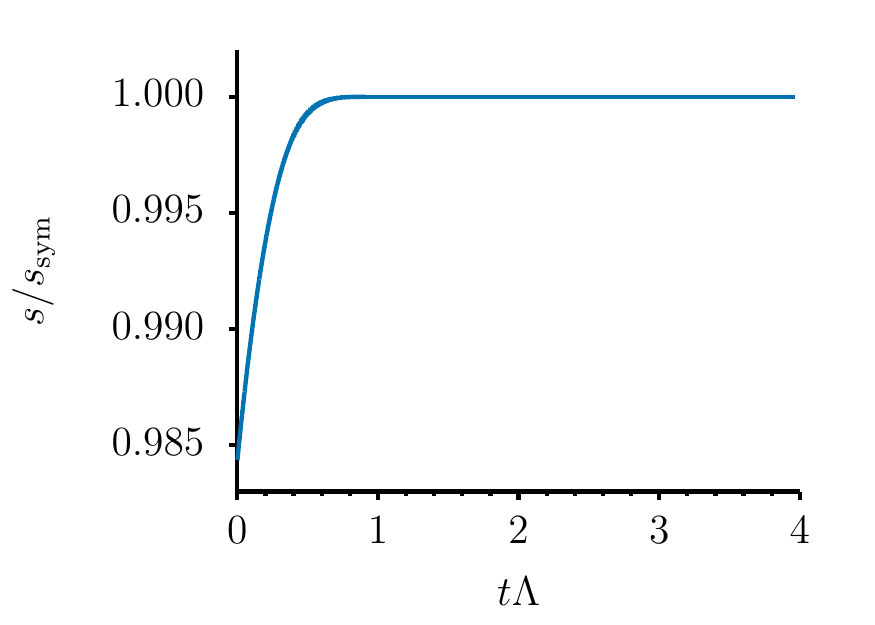} \hfill
\includegraphics[width=2.9in]{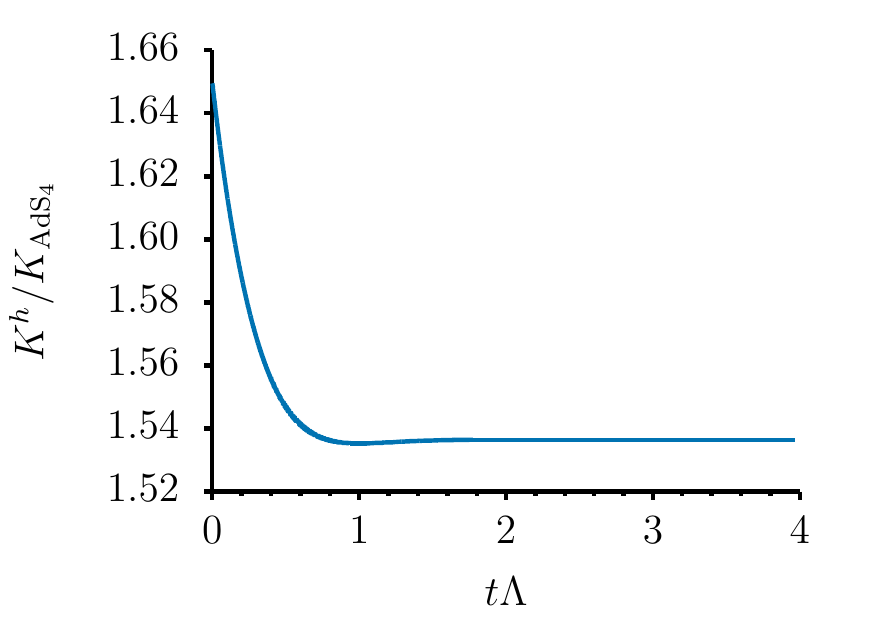}
\end{center}
 \caption{Dynamical entropy density $s$ relative to the equilibrium entropy 
density $s_{sym}$ in $\zet_2$-symmetric sector of exotic black holes (left panel). 
Corresponding evolution of the bulk Kretschmann scalar (right panel).  } \label{figure7}
\end{figure}

Results of a typical evolution are presented in figures~\ref{figure6} and~\ref{figure7}. 
Here, the energy density is below the critical one \eqref{ecrit}, 
\begin{equation}
\cale= 0.793642\ \cale_{crit}\qquad \Longleftrightarrow\qquad  \mu=-4 \Lambda^3
\;.
\eqlabel{esymsim}
\end{equation}
The left panel of figure~\ref{figure6} 
shows time evolution of the expectation value of $\calo_r$. Within a time scale 
$t\sim \Lambda^{-1} $ the system equilibrates. The equilibrium expectation value,
defined as  
\begin{equation}
\langle\calo_r^e\rangle =\lim_{t\Lambda\to \infty} \langle\calo_r(t)\rangle \;,
\eqlabel{defsysequi}
\end{equation}
is represented by a (green) dashed  line. We used the value of  $\langle\calo_r^e\rangle$ 
obtained from the independent analysis of the static configurations, reported in section \ref{static},
evaluated at the energy density \eqref{esymsim}. Consistency of \eqref{defsysequi} is an important check 
of our evolution. The right panel of figure~\ref{figure6} illustrates the system's approach to equilibrium,
which displays a typical $\phi$ quasinormal mode ring-down of the exotic black hole horizon.

The entropy density is an intrinsically equilibrium concept in QFTs. 
One benefit of the holographic framework is that it provides a well-motivated 
notion of the non-equilibrium (even far from equilibrium) entropy. 
Following \cite{Booth:2005qc,Figueras:2009iu} we identify  nonequilibrium 
entropy density $s$ with the Bekenstein-Hawking entropy corresponding to the apparent horizon 
(see \eqref{ldata}) area density
\begin{equation}
s(t)=\frac{2\pi}{\k^2}\ \Sigma(t,r)^2\bigg|_{r=1} \;.
\eqlabel{entsymm}
\end{equation}
The right panel of figure~\ref{figure7} shows the evolution of thus defined  dynamical entropy density. 
Notice that in line with the second law of thermodynamics,  $\dot s(t)\ge 0$ and approaches 
at late times the equilibrium value $s_{sym}$, computed independently for the static configuration 
with the energy density \eqref{esymsim}.

While we can study the non-equilibrium dynamics in $\zet_2$-symmetric sector of the holographic model \eqref{s4},
completely suppressing the $\chi$-scalar fluctuations as in \eqref{pqinitial},
at $\cale<\cale_{crit}$, and in particular in the discussed example \eqref{esymsim}, this is an unphysical approximation ---
in realistic settings the fluctuations of the $\chi$ scalar will always be present, and would destabilize this 
$\zet_2$-symmetric dynamics. In the right panel of figure~\ref{figure7} we show the time dependence of the  
bulk Kretschmann scalar $K$ evaluated at the apparent horizon, 
\begin{equation}
K(t)=R_{abcd}R^{abcd}\bigg|_{(t,r=1)}\;,
\eqlabel{defK}
\end{equation}
relative to the $AdS_4$  Kretschmann scalar $K_{AdS_4}$ (recall $K_{AdS_4}={\rm const}=24$), 
to emphasize the fact that 
even if the symmetric sector is unstable, its bulk dynamics is weakly curved. Thus, 
higher derivative supergravity and string corrections are arguably irrelevant for the 
onset of the $\zet_2$ symmetry breaking instability of the exotic black hole horizons.

\subsection{Long-wavelength (GL-type) instability of the symmetric sector}\label{gltypeinstab}

In this section we study {\em linearized fluctuations} of the $\zet_2$ symmetry breaking operator 
$\calo_i$ in the symmetric phase of the holographic model \eqref{s4}. We initialize 
the symmetric sector of the model as explained in section \ref{dynsym} for energy 
densities above/below the critical one. The bulk scalar $\chi$, dual to an irrelevant 
operator $\calo_i$, is initialized as (in $\l_0=0$ gauge, see Appendix \ref{initial})
\begin{equation}
\chi\bigg|_{t=0}=\cala_q\ \frac{\exp\left(-\frac 1r\right)}{r^4} \;.
\eqlabel{chiinit}
\end{equation} 
To treat symmetry breaking in a linear approximation, we set $\chi(t,x)\equiv 0$ 
in all dynamical equations, except for the third equation in \eqref{evolveoms} --- the only one 
linear in the field $\chi$ and which determines its dynamics --- which is kept unchanged.

\begin{figure}[t]
\begin{center}
\includegraphics[width=2.9in]{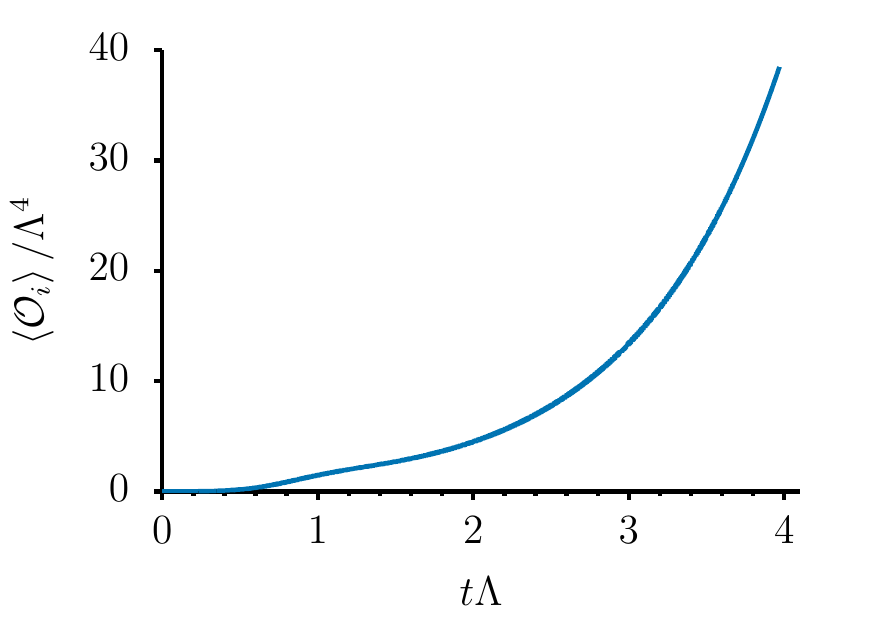} \hfill
\includegraphics[width=2.9in]{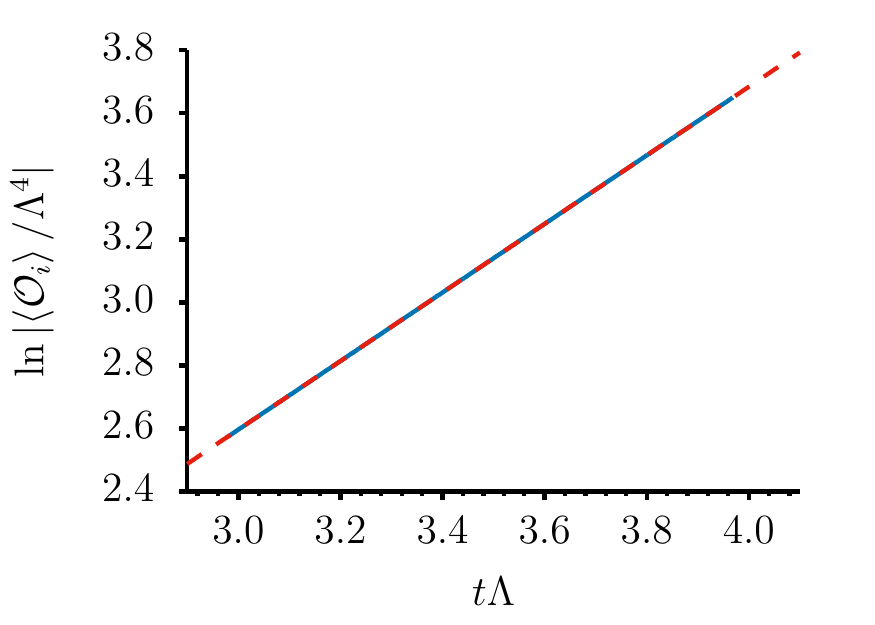}
\end{center}
 \caption{Linearized fluctuations of the symmetry breaking operator $\calo_i$ during the 
dynamical evolution 
of the $\zet_2$-symmetric sector of exotic black holes with $\cale<\cale_{crit}$. The dashed 
red line (right panel) is the linearized fit to the exponential growth of  $\langle\calo_i\rangle$ 
at late times.} \label{figure8}
\end{figure}

Figure~\ref{figure8} presents the linearized fluctuations of the symmetry breaking operator $\calo_i$
during evolution of the symmetric sector with 
\begin{equation}
\cale= 0.793642\ \cale_{crit}\qquad \Longleftrightarrow\qquad  \mu=-4 \Lambda^3
\;.
\eqlabel{esymsim1}
\end{equation}
After a time $t\sim \Lambda^{-1}$ the symmetric sector equilibrates, and $\langle\calo_i\rangle$
exhibits an exponential growth with time as it evolves over such state. 
The growth rate can be extracted at late times [(red) dashed  line,  right panel]:
\begin{equation}
\begin{split}
&\ln|\langle\calo_i\rangle/\Lambda^4|\bigg|_{red\ line\ fit}=-0.66414(3) +
1.0869(7)\ t\Lambda \;, \\
&\Im(\w_\chi)/\Lambda\bigg|_{fit}= 1.0869(7) \;.
\end{split}
\eqlabel{ratebelow}
\end{equation} 
This is in excellent agreement with the independent computation of the $\chi$-scalar QNM frequencies 
reported in figure~\ref{figure1} at energy density \eqref{esymsim1}:
\begin{equation}
\frac{\Im(\w_\chi)\bigg|_{fit}}{\Im(\w_\chi)\bigg|_{QNM}}=0.99997(3) \;.
\eqlabel{check1}
\end{equation}

\begin{figure}[t]
\begin{center}
\includegraphics[width=2.9in]{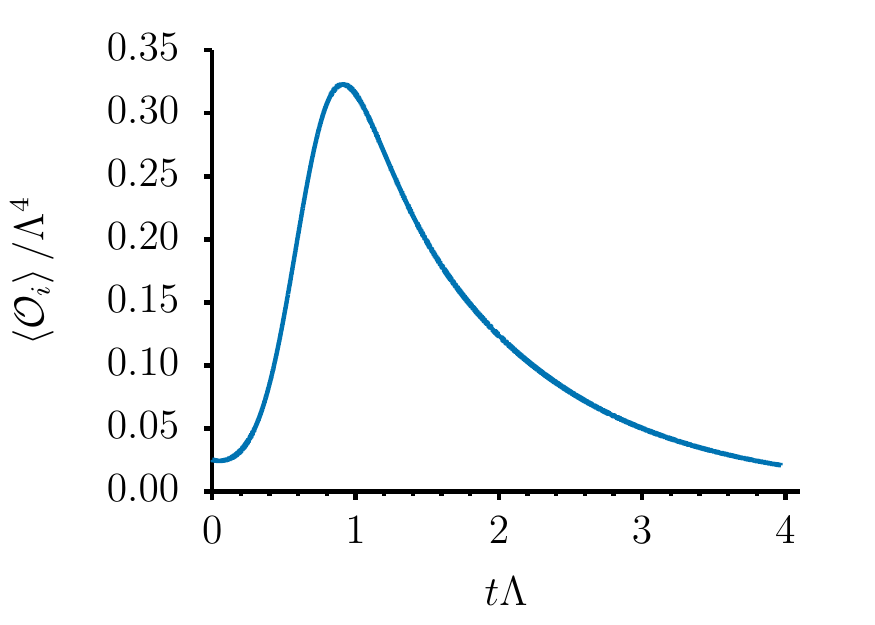} \hfill
\includegraphics[width=2.9in]{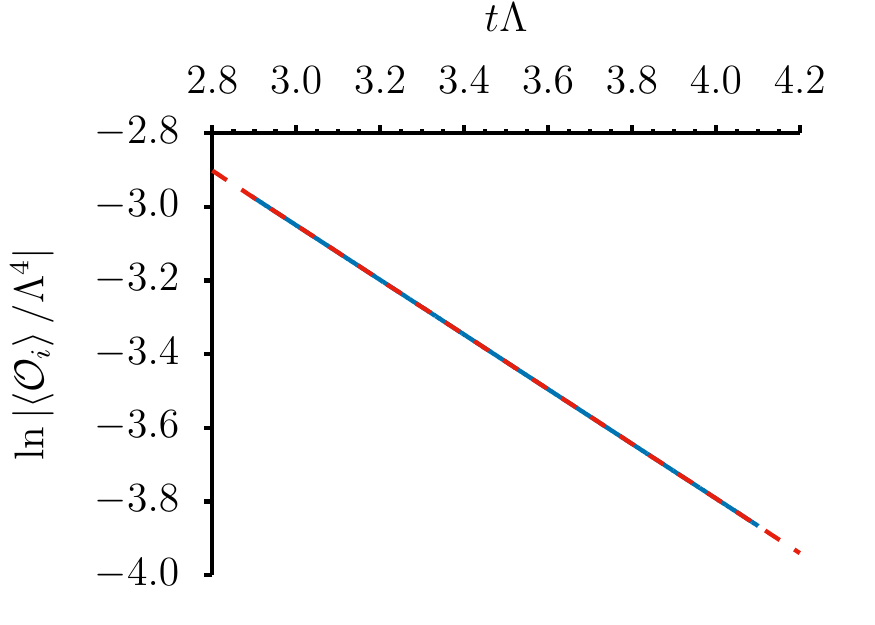}
\end{center}
 \caption{Linearized fluctuations of the symmetry breaking operator $\calo_i$ during dynamical evolution 
of the $\zet_2$-symmetric sector of exotic black holes with $\cale>\cale_{crit}$. The  (red) dashed 
 line (right panel) is the linearized fit to the exponential decay of  $\langle\calo_i\rangle$ 
at late times.} \label{figure9}
\end{figure}

Figure~\ref{figure9} presents the linearized fluctuations of the symmetry breaking operator $\calo_i$
during evolution of the symmetric sector with 
\begin{equation}
\cale=1.1904(6)\ \cale_{crit}\qquad \Longleftrightarrow\qquad  \mu=-6 \Lambda^3
\;.
\eqlabel{esymsim2}
\end{equation}
Again, after a time $t\sim \Lambda^{-1}$ the symmetric sector equilibrates, and the evolution of
$\langle\calo_i\rangle$
over such state exhibits an exponential decay with time. The decay rate can be extracted at late times 
[(red) dashed  line,  right panel]:
\begin{equation}
\begin{split}
&\ln|\langle\calo_i\rangle/\Lambda^4|\bigg|_{red\ line\ fit}=-0.28139(8) -
0.90347(9) \ t\Lambda \;,\\
&\Im(\w_\chi)/\Lambda\bigg|_{fit}=  - 0.90347(9) \;.
\end{split}
\eqlabel{rateabove}
\end{equation} 
This also agrees with the independent computation of the $\chi$-scalar QNM frequencies 
reported in figure~\ref{figure1} at energy density \eqref{esymsim2}:
\begin{equation}
\frac{\Im(\w_\chi)\bigg|_{fit}}{\Im(\w_\chi)\bigg|_{QNM}}=1.0000(2) \;.
\eqlabel{check2}
\end{equation}

Notice that the fluctuations of $\chi$ do not oscillate 
(both in the stable, ie. those that give rise to equilibrium, and unstable scenarios), \ie 
\begin{equation}
\Re(\w_\chi)\bigg|_{QNM}=0 \;.
\eqlabel{noreal}
\end{equation}
We believe this is a reflection of the spontaneous character of the symmetry breaking due to these fluctuations 
at the horizon\footnote{Similar phenomena was observed in \cite{Buchel:2010wp}.} together with the boundary
conditions adopted.

\subsection{Fully non-linear evolutions of  stable and unstable black holes}\label{fullsim}
We now turn our attention to the fully non-linear behavior.
Thanks to the simulations' ability to account for the backreaction of the 
field $\chi$ a rich phenomenology is uncovered.
To aid in the interpretation of the results, we monitor several quantities:

\begin{itemize}
\item The dynamical behavior of $p_2$ and $q_4$.
\item The area of the Apparent and Event horizons (see Appendix~\ref{evhorfinder}).
\item The behavior of the Kretschmann curvature scalar $K= R_{abcd} R^{abcd}$ (normalized
by the value of $K$ for pure AdS).
\end{itemize}

\begin{figure}[t]
\begin{center}
  \includegraphics[width=2.9in]{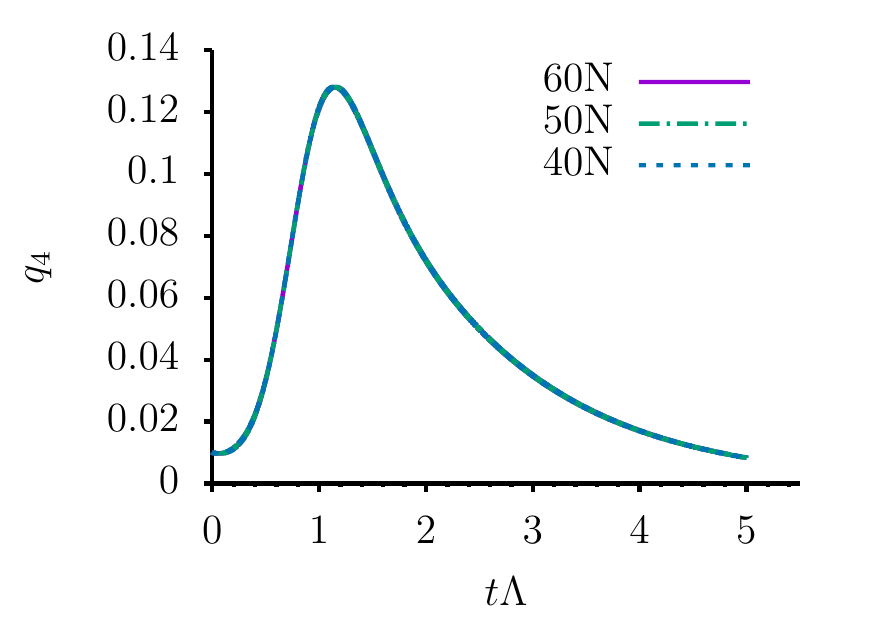} \hfill
  \includegraphics[width=2.9in]{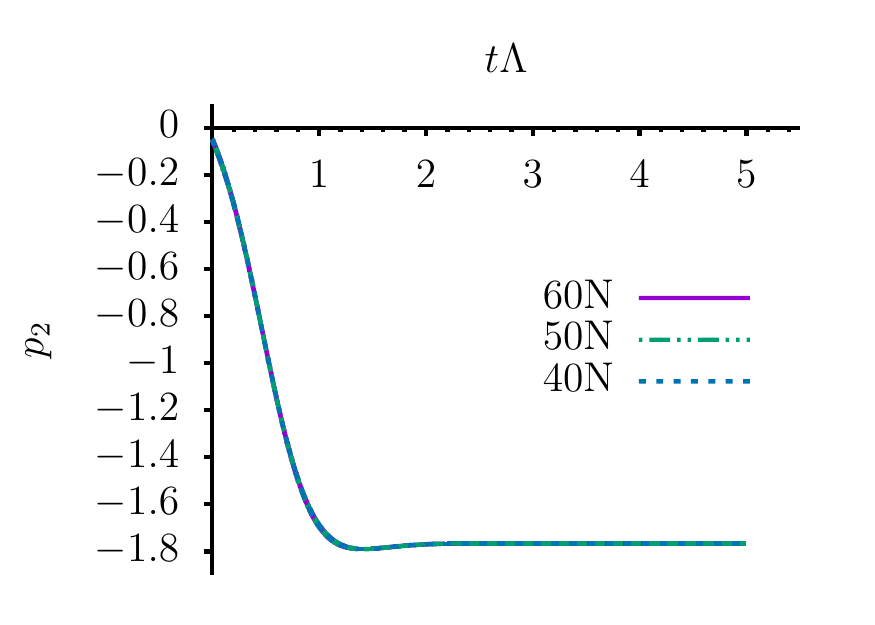}
\end{center}
 \caption{$p_2$ and $q_4$ vs. time for the stable case.} \label{figure_p2q4stable}
\end{figure}

\begin{figure}[t]
\begin{center}
  \includegraphics[width=2.9in]{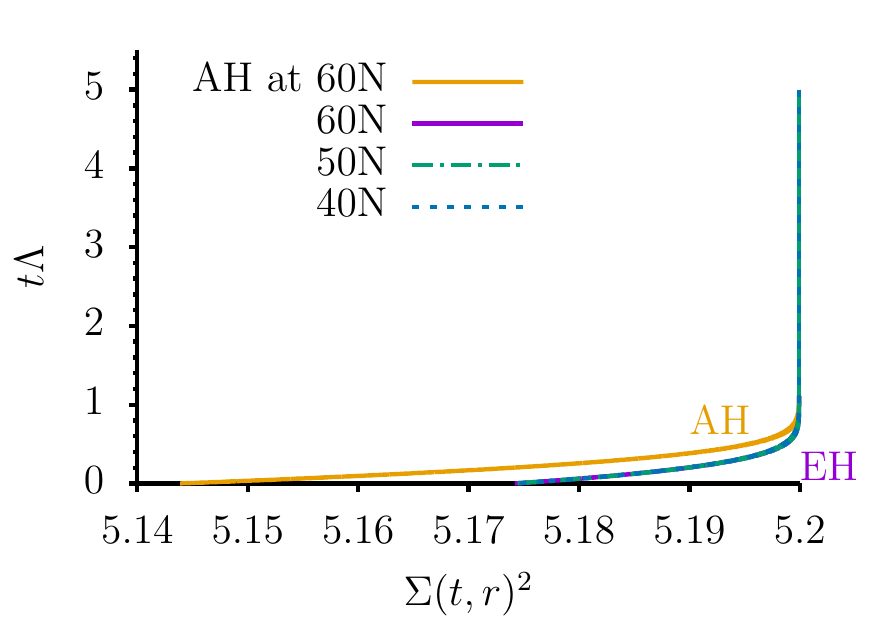} \hfill 
  \includegraphics[width=2.9in]{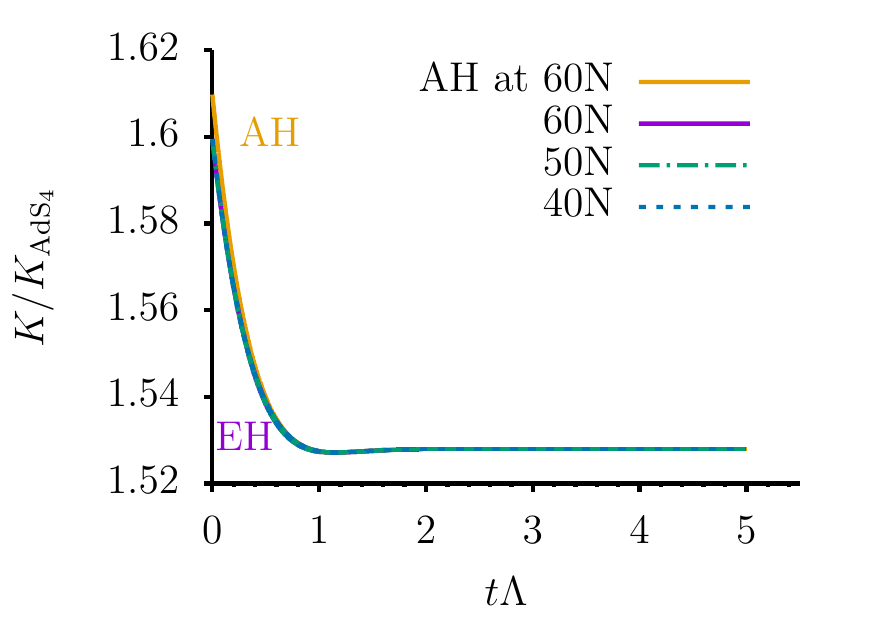}
\end{center}
 \caption{Area and Kretschmann for stable case.} \label{figure_areakretshstable}
\end{figure}

As a first case of study, we confirm that for $\cale>\cale_{crit}$ the behavior observed is consistent
with that captured by the linearized analysis described in section~\ref{gltypeinstab}. For this case,
the system asymptotically approaches a stationary hairy black hole which is evidenced by
a non-zero asymptotic value of $p_2$ as illustrated in figure \ref{figure_p2q4stable} as well
as the behavior of the normalized curvature scalar $K$ shown
in figure \ref{figure_areakretshstable}. This figure, also shows that at late times
the event and apparent horizon coincide and remain stationary.

On the other hand, the case where $\cale<\cale_{crit}$ --identified in the previous section as unstable--
leads to a markedly different behavior. For concreteness, we concentrate
on the particular case defined by the following configuration.
\nxt Energy density (with $\Lambda=1$)
\[
\cale=0.793642\ \cale_{crit}\qquad \Longleftrightarrow\qquad  \mu=-4 \Lambda^3
\;.
\]
\nxt Initial conditions are chosen describing a perturbed black hole
with both non-zero $\phi$ and $\chi$ (as detailed in Appendix \ref{initial}) with 
\[
\cala_p=1.0\,,\qquad \cala_q=0.01 \, .
\]
Under these conditions, the system gives rise to a rich --and very rapidly evolving dynamics--
which we have confirmed through extensive convergent studies. For instance, by inspection
of results obtained with different number of collocation points ($N=20,30,40,50,60$ points), use of
adaptive time-stepping to capture the increasing faster dynamics observed, and employing a different
coordinate condition (setting $\lambda(t)=0$) which does not keep the apparent horizon at a 
fixed location as done in~\cite{Bosch:2016vcp}. All these studies confirm the observed behavior that 
we describe next.

\begin{figure}[t]
\begin{center}
  \includegraphics[width=2.9in]{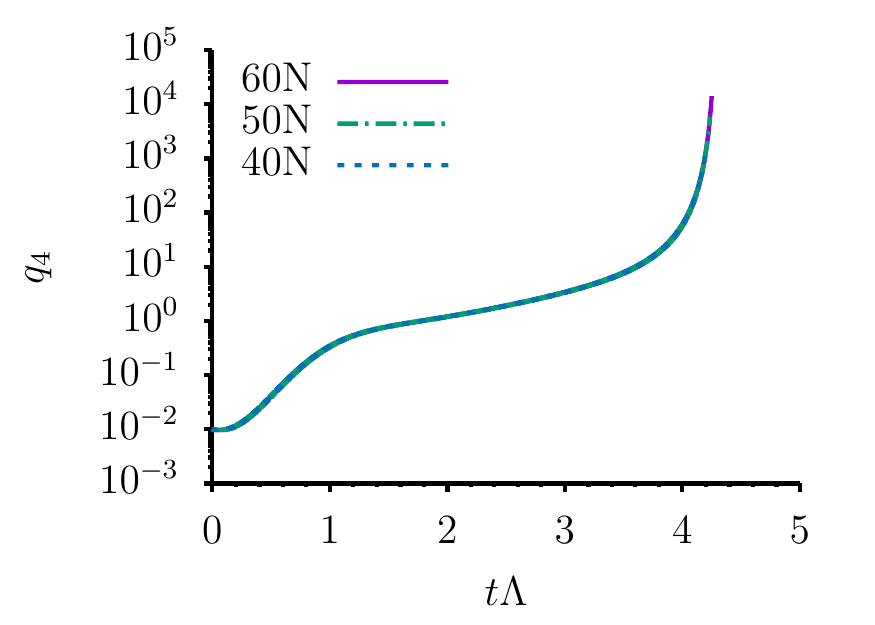} \hfill 
  \includegraphics[width=2.9in]{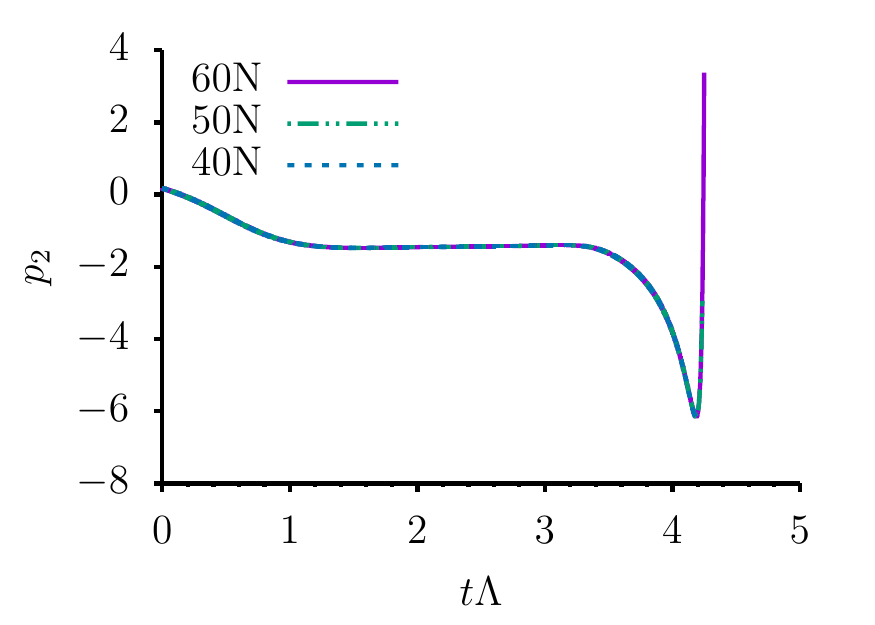}
\end{center}
 \caption{$p_2$ and $q_4$ vs. time for the unstable case.} \label{figure_p2q4unstable}
\end{figure}

\begin{figure}[t]
\begin{center}
  \includegraphics[width=2.9in]{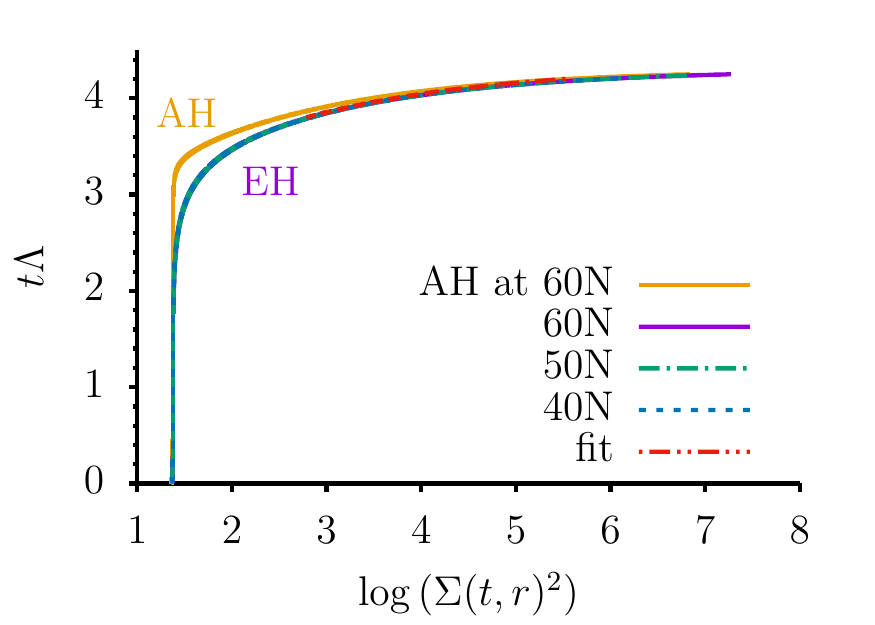}\hfill 
  \includegraphics[width=2.9in]{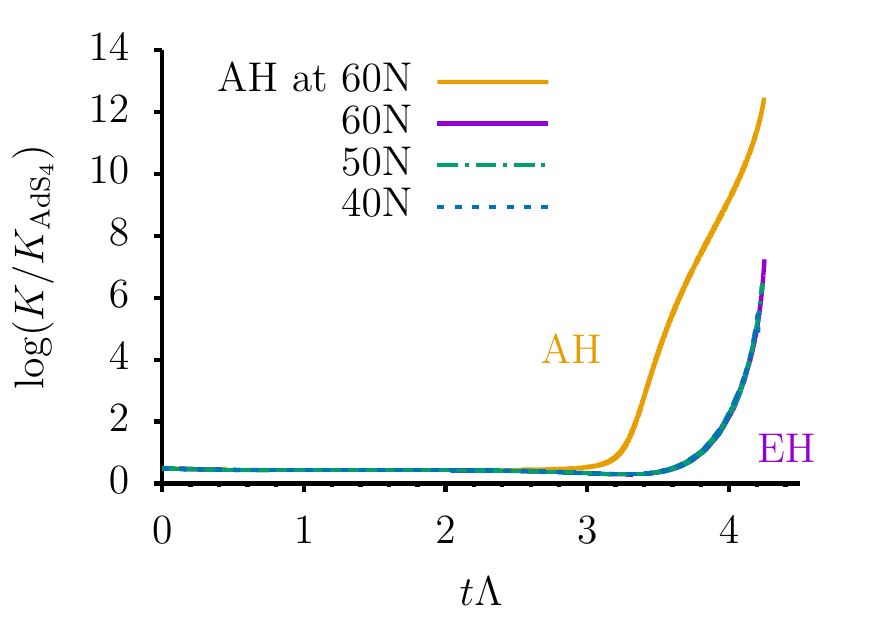}
\end{center}
 \caption{Area and Kretschmann for unstable case.} \label{figure_areakretshunstable}
\end{figure}

\begin{figure}[t]
\begin{center}
  \includegraphics[width=2.9in]{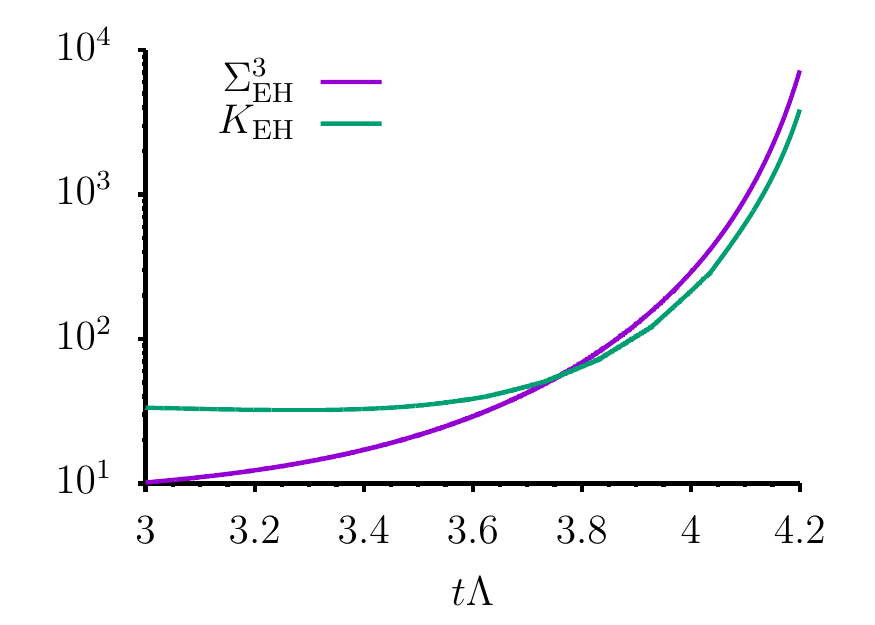}
\end{center}
\caption{Similar asymptotic behaviour of the Kretschmann scalar and $\Sigma^3$
at the event horizon.}
\label{fig:asymp_kret_area}
\end{figure}

\begin{figure}[t]
\begin{center}
  \includegraphics[width=3.2in]{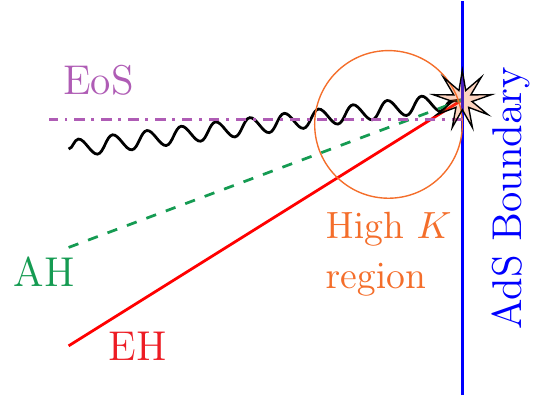}
\end{center}
 \caption{Schematic spacetime diagram. 
As time progresses, the apparent horizon approaches the event horizon in finite
time. The scalar curvature diverges and arbitrarily high curvature regions
can be identified by asymptotic observers at finite times. In the diagram,
``EoS'' refers to the ``End of the Simulation'' while the star refers to the blow up of
the Kretschmann at the boundary in finite time.} \label{spacetimediagram}
\end{figure}

As the field $\phi$ ``rolls down'' the unbounded potential, the energy gained impacts
the dynamics of its normalizable coefficient ($p_2$) as well as the normalizable
coefficient of $\chi$ which grows without bounds as shown in figure \ref{figure_p2q4unstable}. 
This behavior is evidenced in the black hole, which grows fast and eventually reaches the AdS 
boundary in {\em finite asymptotic time} as illustrated in figure~\ref{figure_areakretshunstable}. 
The figure shows both the apparent horizon (AH), and the event horizon (EH) as well as the curvature scalar
evaluated on them. Clearly, as time progresses, the AH approaches the EH and both asymptote to
infinite size in a finite amount of time. This asymptotic behavior can be fit by the expression 
\begin{equation}
\Sigma^2_{\mathrm{EH}} \propto 1/(a + b t + c t^2) \;,
\end{equation}
with the following coefficients $\{a = 3.934(5)$, $b=-1.811(5)$ and
$c=0.2084(5)\}$. This fit indicates a finite time divergence at $t \approx 4.30$. We find 
a similar asymptotic behaviour for the Kretschmann scalar evaluated at the horizon with
$K \simeq \Sigma^3_{\mathrm{EH}}$, as seen in figure~\ref{fig:asymp_kret_area}. Thus, at
late times,  
\begin{equation}
K_{\mathrm{EH}} \propto \Sigma^3_{\mathrm{EH}} \propto \left(\frac{1}{a+bt+ct^2}
\right)^{\frac{3}{2}} \; ;
\end{equation}
consequently, $K_{\mathrm{EH}}$ diverges in finite time at the boundary
of AdS.
Additionally, the  (normalized) scalar curvatures on the AH and EH diverge 
with $K_{\mathrm{EH}} \le K_{\mathrm{AH}}$.
Naturally, the code is eventually  unable to keep up with the radically rapid dynamics which requires
ever smaller timesteps to capture the following behavior. Nevertheless, we have been able to extract
convergent solutions up to a sufficiently late stage to understand the behavior and fate of the spacetime.
The picture that arises is that the spacetime explores arbitrarily large curvatures
in finite time, and outgoing null geodesics emanating from such regions reach the boundary of AdS 
in finite asymptotic time as indicated in figure \ref{spacetimediagram}. 
This behavior would violate the spirit of the weak cosmic censorship conjecture, in that
far observers can be reached by signals emanating from arbitrarily curved spacetime regions, and is similar to that 
recently reported in~\cite{Crisford:2017zpi}.

\section{Conclusions}\label{end}

In this work we studied an interesting instability of the black hole horizons, observed first 
in \cite{Buchel:2009ge}: 
below some critical energy density the horizon is unstable with respect to fluctuations 
spontaneously breaking a discrete symmetry. However, there is no static end-point 
associated with the nonlinear build-up of the  symmetry breaking condensate (scalar hair at 
the horizon), as opposed to typical constructions of holographic superconductors \cite{Hartnoll:2008vx}.
The instability is perturbative in nature (\ie describing a  second order phase transition), 
and is triggered by an arbitrary small amplitude 
of the symmetry breaking mode, provided the conserved energy density of the state $\cale$ 
is below a critical energy density. As a result, the onset of the instability, and dynamics 
close to it, can not be affected by higher-order nonlinearities in the gravitational scalar 
potential as long as the amplitude of unstable modes remain small. 
Additionally, the instability initiates in the long wavelength regime,
\ie at small bulk curvature, and thus can not be removed by higher derivative corrections 
to the gravitational effective action. 

We studied the future development of the instability,
using a characteristic formulation of asymptotically anti-de Sitter gravitational dynamics     
\cite{Chesler:2013lia}, and argued that, at the classical level, 
the end point of the instability induces a curvature singularity at finite asymptotic time.
Specifically, we demonstrated an apparently unbounded growth of the Kretschmann scalar 
 in the bulk  (\eg evaluated at the location of the apparent horizon). 
 Thus, our model \eqref{s4} provides a simple example of arbitrarily large curvatures
arising in asymptotically anti-de Sitter space times at sufficiently late times.

While the analysis of this work is focused on a specific phenomenological model of gauge theory/gravity correspondence, 
represented by the effective gravitational action \eqref{s4}, the phenomena described, \ie horizon 
instability without the static end point, is realized within bona fide holographic correspondence 
scenarios (\eg~\cite{Donos:2011ut,Dias:2015pda}). We expect that curvature singularities 
also arise in those models 
as well\footnote{It is a straightforward exercise to examine this in a holographic model \cite{Donos:2011ut};
dynamics of small black hole localization in $AdS_5\times S^5$ is much mode difficult \cite{bcy}. }. 
The observation that curvature singularities might arise dynamically implies that consistent truncations 
of string theory and supergravity, 
while suitable to address static states in the theory, may fail in dynamical settings ---    
when the evolution enters the regime of highly curved geometry; and in such cases 
stringy corrections will be important.

What are the holographic implications of a singularity developing evolution for the 
boundary gauge theory? A standard lore is that states of a closed non-integrable interacting 
system with large number of degrees of freedom should dynamically equilibrate \cite{ms}. 
In the context of holography, early indications supported this 
for generic far-from-equilibrium, arbitrary low-energy states of strongly coupled conformal 
gauge theories, even when the dynamical evolution of these states was artificially 
restricted to symmetric submanifolds of the full phase space of the 
theory\footnote{The states in question were spatially isotropic, and 
invariant under all global symmetries, \ie the R-symmetry.} 
\cite{Bizon:2011gg,Buchel:2012uh}. Shortly thereafter it was argued 
\cite{Buchel:2013uba,Balasubramanian:2014cja,Green:2015dsa}
that in fact symmetric phase space of holographic conformal field theories
has islands of stability that never equilibrate. In this study we identified yet another possibility:
initial states of holographic strongly coupled gauge theories, well described classically 
in the gravitational frame, evolve to a singularity in finite time. 
Singularity is a signature for a breakdown of an 
approximation, and we see two possible reasons. First, it is possible the singularity is an artifact 
of our restriction of the state evolution to symmetric submanifolds of the full phase space
of the theory and that symmetry breaking modes would allow for a smooth evolution. 
As we discuss below, hydrodynamic modes in the system can be gapped, leading to the same qualitative behaviour. 
It is more difficult to argue for the absence of light modes spontaneously breaking   
internal symmetries --- the singularity observed  might be an indication that some of these symmetries 
must be dynamically broken during the evolution (similar ideas were proposed in \cite{Dias:2016eto,bcy}). 
Second, the state evolution in the gravitational frame of the holography might not be always semiclassical.

We find it important to discuss another possible limitation of the study carried out here
and its conclusions. 
We restricted the dynamics in our model to preserve boundary homogeneity and isotropy. 
One might argue that the physical phenomena discussed here arise as a consequence of such a restriction,
and that a sufficiently generic initial state would smoothly evolve to an end point where
these symmetries are spontaneously broken. We do not have a full answer to this question --- gravitational simulations 
in the presence of spatial inhomogeneities are beyond the scope of this paper.
At the very least, in the holographic model studies  in \cite{Withers:2013loa} there is an exotic instability discussed,
without the  spatially modulated endpoint as well\footnote{We would like to thank Ben Withers 
for bringing the reference to our attention.}.
Since the instability and the evolution towards the singularity in our model can be triggered by 
arbitrarily small amplitude fluctuations of the $\chi$-mode, \ie energetically arbitrary close to 
the critical point, the potentially physics-modifying hydrodynamics modes can be gapped, 
rendering them irrelevant to the question as to whether or not the singularity observed is physical.   
To demonstrate this, we modified our model with a boundary with topology $R^{2}\to S^2$. 
All the main features described in the former case remain in the latter, in particular: 
there is a horizon instability,  there is no static end point 
below some critical energy density associated with the onset of the perturbative instability, 
the low-energy $SO(3)$-invariant states evolve to a singular solution. 
It would be interesting to explore in details  the role of additional massless fields
at the threshold of instability, and their effect on the singularity development.

A feature of the bulk scalar potential of our holographic model  \eqref{s4} 
is that it is unbounded in the $\chi$-direction (recall that the nonlinear coupling $g<0$).
We study in Appendix \ref{secbounded} a modification of the model which ``bounds'' the $\chi$-potential 
with higher order, nonlinear in $\chi$, interactions\footnote{We would like to thank Jorge Santos
for raising the issue of the unboundedness of the scalar potential in our model with its potential effect on the 
singular evolution  that prompted this analysis.}. Of course, the linearized instability
is unaffected; likewise, the unstable phase with $\langle \calo_i\rangle\ne 0$ 
for $\cale>\cale_{crit}$ is unchanged qualitatively (close to $\cale_{crit}$ the higher-order nonlinear terms in the 
gravitational potential are suppressed). However, we find a new static black hole phase
with $\langle \calo_i\rangle\ne 0$ , that exists 
both for $\cale<\cale_{crit}$ and $\cale>\cale_{crit}$ for the bounded potentials and, at least in the vicinity of 
perturbative instability, has higher entropy density than the symmetric phase. Unlike the exotic branch of the 
black holes, this new phase does not bifurcate from the onset of long-wavelength
 instability of the symmetric phase.
For $\cale<\cale_{crit}$ this new static phase is always the end point of the evolution; for $\cale>\cale_{crit}$
the new symmetry broken phase can only be reached if the initial amplitude of the symmetry breaking fluctuations is 
sufficiently large --- the symmetry broken phase is a potential barrier separated from the symmetric phase 
whenever $\cale>\cale_{crit}$. Our model 
\eqref{s4} is a phenomenological example of the holographic correspondence, thus one might worry whether 
curvature diverging scenarios described here is realized in genuine (top-down) holographic dualities. 
We believe the answer to the question is in the affirmative:
\nxt  First, the unbounded potentials are ubiquitous in holography --- a typical example is
a well-studied  $\caln=2^*$ holography  \cite{Pilch:2000ue,Buchel:2000cn,Buchel:2013id},
where the bulk gravitational scalars $\{\alpha,\chi\}$ effective action takes form,
\begin{equation}
\begin{split}
&S_{\caln=2^*}^{scalar}\ \sim\ \frac{1}{16 \pi G_{5}} \int \mathrm{d}^{5} x \sqrt{- g} \biggl( 
-12 (\partial \alpha)^{2} - 4 (\partial \chi)^{2} - V 
 \biggr) \, , \\
&V(\alpha,\chi) =   -  e^{-4 \alpha} - 2 e^{2 \alpha} \cosh{2 \chi} + \frac 14 e^{8 \alpha} \sinh^{2}{2 \chi} \, .
\end{split}
\eqlabel{n2potential} 
\end{equation}
The reason why the scalar potentials in supergravity constructions can be unbounded comes 
from the fact that they arise from the superpotential as (for the $\caln=2^*$ example
\eqref{n2potential})
\begin{equation}
\begin{split}
&V=\frac{1}{16}\left[\frac 13 \left(\frac{\del W}{\del
\a}\right)^2+ \left(\frac{\del W}{\del \chi}\right)^2\right]-\frac
13 W^2\, , \\
&W=- e^{-2\alpha} - \frac{1}{2} e^{4\alpha} \cosh(2\chi) \, .
\end{split}
\eqlabel{superpotential}       
 \end{equation}
It is the $-\frac 13 W^2$ contribution to $V$ in \eqref{superpotential} that is responsible for the 
unboundedness of $V$. 
\nxt Second, the scalar potential in the top-down embedding of the exotic black hole phenomena 
\cite{Buchel:2009ge} constructed in  \cite{Donos:2011ut} (see eq.(2.32) there) is unbounded from below:
\begin{equation}
V(\varphi)=-2\ \left(2 +\cosh(2\varphi) \right) \, .
\eqlabel{dg}
\end{equation}

There are lots of open questions left for the future. 
It would be interesting to understand whether the divergent curvature scenario proposed here is universal. 
Is it possible to understand analytically the approach towards the singularity as in explorations
of the BKL conjecture? (e.g.~\cite{Garfinkle:2003bb}).
 The link between the boundedness 
of the gravitational potential and the singular evolution should be studied in more detail. 
It is certainly important to understand the consequences of the diverging curvature for the boundary gauge theory. 
Is there a QFT-solvable holographic example that captures the proposed singular behavior?


\section*{Acknowledgments}
Research at Perimeter
Institute is supported by the Government of Canada through Industry
Canada and by the Province of Ontario through the Ministry of
Research \& Innovation. This work was further supported by
NSERC through the Discovery Grants program (AB and LL), by CIFAR (LL) and by
CONACyT-Mexico (PB).

\appendix
\section{Appendix: Numerical setup}\label{num}

We adapt the characteristic formulation of \cite{Chesler:2013lia} for the numerical 
solution of \eqref{evolveoms}-\eqref{coneoms2}.

\subsection{Field redefinitions and the code equations}

We introduce a new radial coordinate  
\begin{equation}
x\equiv \frac 1r\in [0,1]\,,\qquad d_+ =\del_t +A(t,r)\ \del_r \ \to\ \del_t-x^2 A(t,x) \ \del_x\,,  
\eqlabel{redef}
\end{equation}
maintaining $'\equiv \del_x$ and $\dot\ \equiv \del_t$,
and redefine the fields 
\begin{equation}
\{\, \phi\,,\, \chi\,,\, \Sigma\,,\, A\,,\, d_+\phi\,,\, d_+\chi\,,\, d_+\Sigma\, \}
\ \to\
\{\, p\,,\, q\,,\, \sigma\,,\, a\,,\, dp\,,\, dq\,,\, d\sigma\, \}
\eqlabel{fieldsor}
\end{equation}
as follows
\begin{equation}
\begin{split}
&\phi(t,x)=x\ p_1+x\ p(t,x)\;,\\
&\chi(t,x)=x^3\ q(t,x)\;,\\
&\Sigma(t,x)=\frac 1x+\sigma(t,x)\;,\\
&A=a(t,x)+\frac 12\ \Sigma(t,x)^2\;,\\
&d_+\phi(t,x)=-\frac{p_1}{2}+x\ dp(t,x)\;,\\
&d_+\chi(t,x)=x^3\ dq(t,x)\;,\\
&d_+\Sigma(t,x)=x\ \ds(t,x)+\frac 12\ \Sigma(t,x)^2-\frac{p_1}{12}\
d_+\phi(t,x)+\frac{p_1^2}{48} \;.
\end{split}
\eqlabel{fieldnew}
\end{equation}
Using \eqref{basym}, we find the asymptotic boundary expansion $x\to 0_+$ for the new fields:
\begin{equation}
\begin{split}
&p=p_2(t)\ x + \calo(x^2)\,,\qquad q=q_4(t)\ x+\calo(x^2)\;,\\
&dp=-p_2(t)-p_1 \l(t)+\calo(x)\,,\qquad dq=-2 q_4(t)+\calo(x)\;,\\
&\sigma=\l(t)-\frac{p_1^2}{8}\ x+\calo(x^2)\,,\qquad \ds=\mu+\calo(x)\;,\\
&a=-\dot\l(t)+\left(\mu-\frac{p_1^2}{12}\ \l(t)-\frac{p_1}{12}\ p_2(t)\right)\
x+\calo(x^2)\;.
\end{split}
\eqlabel{newbasym}
\end{equation}
In new variables \eqref{fieldnew}, the equations of motion used to evolve the system take form:
\begin{equation}
\begin{split}
&\left[\del_{xx}^2+\frac 2x\ \del_x +\frac{x^4}{4}(3q+x
q')^2+\frac{1}{4}(p_1+p+x p')^2\right]\sigma=J_\sigma\;,\\
&J_\sigma\{p,p',q,q'\}=-\frac{x^3}{4}\left(3q+x
q'\right)^2-\frac{p'}{4}(xp'+2p+2p_1)-\frac{1}{4x}(p+p_1)^2\;,
\end{split}
\eqlabel{ceq1}
\end{equation}
\begin{equation}
\begin{split}
&\left[\del_x+\frac{12\s +12 x \s'-x p_1 (p_1+p+x p')}{12(1+x\s)}\right]\
dp+\left[\frac{x(p_1+p+xp')}{1+x\s}\right]\ \ds= J_2\;,\\
&\left[\del_x+\frac{12\s +12 x \s'+x p_1 (p_1+p+x p')}{12(1+x\s)}\right]\ \ds
+\left[-\frac{x p_1^2(p_1+p+x p') }{144(1+x\s)}\right]\ dp= J_3\;,\\
&J_2\{p,p',q,\sigma,\sigma'\}=\frac{1}{1+x\s}\biggl(-\frac{1}{16} p' (p_1^2+8 \s^2) x
-\frac{1}{16} (p_1^2+8 \s^2) (p+p_1)+\frac12 p_1 \s'\\
& -p' \s-\frac{p'}{2x} +\frac{p}{2x^2}\biggr)-(p+p_1) q^2 g x^4\;,\\
&J_3\{p,p',q,\sigma,\sigma'\}=\frac{1}{1+x\s}\biggl(-\frac{p' p_1}{192}  (p_1^2+8 \s^2) x
-\frac{p+p_1}{192} (48 p \s^2+p_1^3+56 p_1 \s^2)
-\frac{\s'}{48} (p_1^2\\
&+72 \s^2)-\frac{p_1 p' \s}{12}+
\frac 1x \left(-\frac\s2  (p+p_1)^2-3 \s \s'-\frac{p' p_1}{24}\right)
+\frac{1}{x^2}\biggl(-\frac{p^2}{4}
-\frac{11}{24} p p_1-\frac{3}{16} p_1^2
\\
&-\frac32 \s'\biggr)\biggr)+\frac g4 \s  (p+p_1)^2 q^2 x^5
+\frac{g}{12} (p+p_1) (3 p+2 p_1) q^2 x^4
+\frac12 \s q^2 x^3+\frac12 q^2 x^2 \;,
\end{split}
\eqlabel{ceq23}
\end{equation}
\begin{equation}
\begin{split}
&\left[\del_x+\frac{2+3 x\s+x^2\s'}{x(1+x\s)}\right]dq+\left[-\frac{x^2 p_1 q'+3 x p_1 q}{12(1+x\s)}\right]dp
+\left[\frac{x^2q'+3 x q}{1+x\s}\right]\ds=J_4\;,\\
&J_4\{p,q,q',\sigma\}=\frac{1}{1+x\s}\biggl(-\left(q \s g (p+p_1)^2+\frac12 \s^2 q'+\frac{1}{16} p_1^2 q'\right) x
-\frac{q}{16}  \biggl(16 g (p+p_1)^2\\
&\qquad\qquad+3 p_1^2+24 \s^2\biggr)-\s q'+\frac 1x \left(-5 \s q-\frac12 q'\right)
-\frac{7q}{2x^2}\biggr) \;,
\end{split}
\eqlabel{ceq4}
\end{equation}
\begin{equation}
\begin{split}
&\left[\del^2_{xx}+\frac 2x\ \del_x\right]a+\biggl[\frac{1}{(1+x s)^2}
\biggl(-\frac12 p' \s^2 x^2+\left(-\frac12 \s^2 (p+p_1)-p' \s-\frac16 p_1 \s'\right) x\\
&-\s (p+p_1)-\frac{p'}{2} -\frac{3p+2p_1}{6x}\biggr)\biggr]dp+\biggl[\frac{2(x^2 \s'-1)}{x(1+x s)^2}
\biggr] \ds+\left[-\frac{x^3(x q'+3 q)}{2}\right]dq=J_5\;,\\
&J_5\{p,p',q,q',\sigma,\s'\}=\frac{1}{(1+x s)^2}\biggl(
\frac{\s^4}{4} (p')^2  x^4+\frac{\s^3}{2} p'  (p \s+p_1 \s+2 p') x^3
+\biggl(\frac14 \s^4 (p+p_1)^2\\
&+2 p' \s^3 (p+p_1)+\frac32 (p')^2 \s^2-(\s')^2 \s^2\biggr) x^2
+\biggl(\s^3 (p+p_1)^2
+\frac14 p' \s^2 (12 p+11 p_1)+(p')^2 \s\\
&-2 (\s')^2 \s\biggr) x
+\frac14 \s^2 (p+p_1) (6 p+5 p_1)
+\frac\s2  p' (4 p+3 p_1)+\frac18 \s' (-p_1^2+8 \s^2)+\frac14 (p')^2\\
&-(\s')^2
+\frac 1x\biggl(\frac\s2  (p+p_1) (2 p+p_1)
+\frac14 p' (2 p+p_1)+2 \s \s'\biggr)
\\&+\frac{1}{x^2}\biggl(\frac{p^2}{4}+\frac{pp_1}{4}+\frac{p_1^2}{8}+\s'\biggr)
\biggr)
+\frac14 x^2 (q' x+3 q)^2 (\s x+1)^2 \;,
\end{split}
\eqlabel{ceq5}
\end{equation}
\begin{equation}
\begin{split}
&\dot p=dp+\frac12 p' (\s^2+2 a) x^2+\left(\frac12 (p+p_1) (\s^2+2a)+p' \s\right) x+(p+p_1) \s
+\frac{ p'}{2}+\frac{p}{2x}\;,\\
&\dot q=dq+\frac12 q' (\s^2+2 a) x^2+\left(\frac32 (\s^2+2a) q+\s q'\right) x+3
\s q+\frac{q'}{2}+\frac{3q}{2x} \;.
\end{split}
\eqlabel{cev}
\end{equation}
Numerical code is organized as follows.
\begin{itemize}
\item {\bf [Step 1]:} assume that at a time step $t$ we have profiles
\begin{equation}
\{p(t,x)\,,\, q(t,x)\,,\, p'(t,x)\,,\, q'(t,x) \}\qquad {\rm and}\qquad \l(t)
\;.
\eqlabel{nstep1}
\end{equation}
\item  {\bf [Step 2]:} we solve linear in $\s$ equation \eqref{ceq1}, subject to boundary conditions
\begin{equation}
\s(t,x=0)=\l(t)\,,\qquad \s'(t,x=0)=-\frac{p_1^2}{8} \;.
\eqlabel{nstep2}
\end{equation}  
\item  {\bf [Step 3]:} we 
solve linear in $\{dp,\ds\}$ system \eqref{ceq23}, subject to the boundary conditions
\begin{equation}
dp(t,x=0)=-p'(t,x=0)-\l(t)\ p_1\,,\qquad \ds(t,x=0)=\mu \;.
\eqlabel{nstep3}
\end{equation}  
\item  {\bf [Step 4]:} we 
solve linear in $dq$ equation \eqref{ceq4}, subject to the boundary conditions
\begin{equation}
dq(t,x=0)=-2 q'(t,x=0) \;.
\eqlabel{nstep4}
\end{equation}  
\item  {\bf [Step 5]:} we 
solve linear in $a$ equation \eqref{ceq5}, subject to the boundary conditions
\begin{equation}
a'(t,x=0)=\mu-\frac{p_1^2}{12}\ \l(t)-\frac{p_1}{12}\ p'(t,x=0)\,,\qquad
a(t,x=1)=a^h \;.
\eqlabel{nstep5}
\end{equation}  
The value $a^h$ is determined from the stationarity of the apparent horizon 
at $x=1$ as explained in the following subsection.
\item  {\bf [Step 6]:} we use evolution equations \eqref{cev}, along with (see \eqref{newbasym})
\begin{equation}
\dot\l(t)=-a(t,x=0) \;,
\eqlabel{defdotl}
\end{equation}
to compute 
\begin{equation}
\{p(t+dt,x)\,,\, q(t+dt,x)\,,\,  \l(t+dt) \;.
\eqlabel{nstep6}
\end{equation}
After computing the radial coordinate derivatives $\{p'(t+dt,x)\,,\, q'(t+dt,x)\}$, we 
repeat {\bf [Step 1]}.
\end{itemize}

Notice that the first equation in \eqref{evolveoms} is redundant in our numerical procedure:
rather than propagating in time $\Sigma$, we compute it from the constraint \eqref{coneoms1} at 
each time step; nonetheless, we monitor the consistency of that equation during the evolution.  

Implementing the code\footnote{Code implementation
is similar to the one used in \cite{Buchel:2014gta}.}, 
we use spectral methods for the radial coordinate integration,
{\bf [Step 2]}- {\bf [Step 5]}. Singularities of the equations at the boundary collocation point 
$x=0$ are resolved using the corresponding boundary conditions instead. We use fourth-order 
Runge-Kutta method for the time evolution, {\bf [Step 6]}.

\subsection{Apparent horizon and the boundary condition for $a$}

Our numerical implementation requires an independent computation of $a^h\equiv a(t,x=1)$ (see \eqref{nstep5}),
given radial profiles $\{p,p',q,q',s,s',dp,ds,dq\}$ and the diffeomorphisms parameter $\l$ at time $t$. 
Following \cite{Chesler:2013lia}, this is done by enforcing the time-independent location of the horizon.
Apparent horizon is located as $x=x_h$ such that 
\begin{equation}
d_+\Sigma(t,x)\bigg|_{x=x_h} =0 \;.
\eqlabel{lochor2}
\end{equation}
Assuming $x_h=1$, $\frac{dx_h}{dt}=0$, and using equations of motion \eqref{ceq1}-\eqref{cev} we compute $a^h$ from
\begin{equation}
\del_t d_+\Sigma(t,x_h)\bigg|_{x_h=1}=0 \;.
\eqlabel{lochor3} 
\end{equation}
Denoting 
\begin{equation}
\left\{\, p^h\,,\, dp^h\,,\, q^h\,,\, dq^h\,,\, \s^h\, \right\}\  \equiv \left\{\, p\,,\, dp\,,\, q\,,\, dq\,,\, \s\, \right\}\bigg|_{(t,x=1)}
\eqlabel{defh}
\end{equation}
we find
\begin{equation}
\begin{split}
a^h=&\frac{1}{4}\biggl(\left(g \left(q^h\right)^2-1\right) \left(p_1+p^h\right)^2+2 \left(q^h\right)^2-6\biggr)^{-1} 
\biggl(\left(2 dp^h-p_1\right)^2+4 \left(dq^h\right)^2\\
&+2 \left(\left(p^h+p_1\right)^2-2 \left(q^h\right)^2+6\right) \left(\s^h+1\right)^2-2 \left(q^h\right)^2 
\left(\s^h+1\right)^2 \left(p_1+p^h\right)^2 g\biggr) \;.
\end{split}
\eqlabel{ahres}
\end{equation}

\subsection{Initial conditions}\label{initial}

To evolve \eqref{ceq1}-\eqref{cev} one has to provide data, at $t=0$ 
as required by {\bf [Step 1]}, see \eqref{nstep1}. In particular, 
we need to specify $\l_0\equiv \l(t=0)$. Once again, we follow \cite{Chesler:2013lia}. 

Recall that both $\phi$ and $\chi$ are left invariant under the reparametrization transformations:
\begin{equation}
\frac 1x\ \to \frac 1x +\l_0 \;.
\eqlabel{repx}
\end{equation}
To maintain this invariance we specify initial conditions 
for $\{p,q\}$ (in $\l_0$-invariant way) in terms of 
two amplitudes $\{\cala_p,\cala_q\}$:
\begin{equation}
\begin{split}
&p\bigg|_{t=0}=\cala_p\ \frac{x}{(1+x\l_0)^2}\
\exp\left[-\frac{x}{1+x\l_0}\right]-\frac{p_1\l_0 x}{1+x\l_0} \;,\\
&q\bigg|_{t=0}=\cala_q\ \frac{x}{(1+x\l_0)^4}\
\exp\left[-\frac{x}{1+x\l_0}\right] \;.
\end{split}
\eqlabel{initcond}
\end{equation}
We then proceed as follows\footnote{For this procedure the integration range over the radial coordinate $x$ 
might exceed unity.}:
\begin{itemize}
\item given  $\{\cala_p,\cala_q\}$ we set $\l_0=0$ and perform {\bf [Step 2]} \eqref{nstep2}
and    {\bf [Step 3]} \eqref{nstep3};
\item having enough data, we follow \eqref{fieldnew} to compute the profile $d_+\Sigma(t=0,x)$;
\item we find numerically the root $x=x_0$ of the equation
\begin{equation}
d_+\Sigma(t=0,x)\bigg|_{x=x_0}=0 \;;
\eqlabel{findx0}
\end{equation}
\item we set the trial value of $\l_0$ as
\begin{equation}
\l_0=\frac{1}{x_0}-1 \;,
\eqlabel{trial}
\end{equation}
which (apart from the numerical errors) would guarantee that the corresponding location of the 
apparent horizon  is now at $x=1$; 
\item the trial value \eqref{trial} is further adjusted repeatedly performing  {\bf [Step 2]} 
and    {\bf [Step 3]} to achieve  
\begin{equation}
d_+\Sigma(t=0,x)\bigg|_{x=1}=0
\eqlabel{adjustl}
\end{equation}
at a high accuracy.  
\end{itemize}

\subsection{Convergence tests}

\begin{figure}[t]
\begin{center}
\includegraphics[width=2.9in]{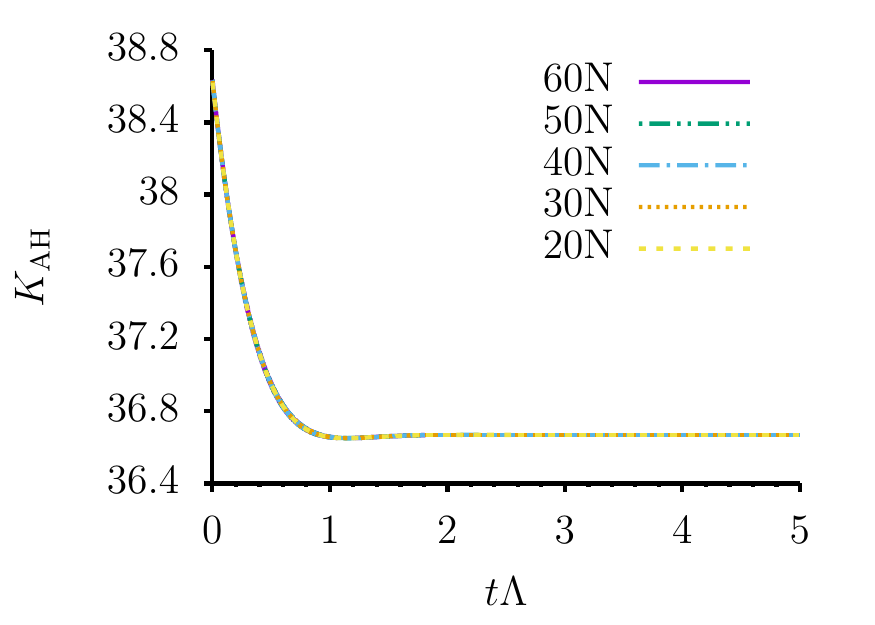} \hfill
\includegraphics[width=2.9in]{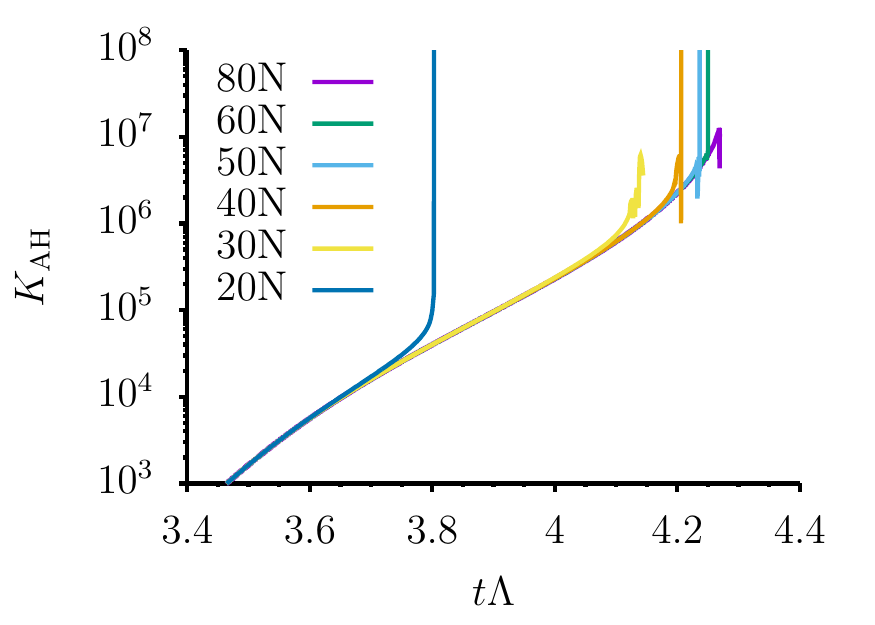}
\caption{Kretschmann scalar at apparent horizon. Left panel stable case. Right
panel unstable case.}
\label{fig:kretsch_convergence}
\end{center}
\end{figure}

We performed self-convergence tests to verify the validity of 
the obtained numerical solutions. In particular, we study each configuration
numerically under different number of collocation points $N=20, 30, 40, 50, 60, 80$. 
We monitored the convergence of the residuals of the constraint equations to zero as well as
each evolved field (and computing self-convergence test by a suitable
interpolation onto a finite difference grid). Additionally, we confirmed convergence of
the location of the event horizon and the Kretschmann scalar at both the apparent and event horizons.
As an illustration, figure~\ref{fig:kretsch_convergence} displays $K_{\rm AH}$ for both the stable
and unstable configurations. For the former case, all resolutions show an excellent agreement.
In contrast, the unstable case illustrates a convergence to a divergent behavior which requires
increasingly finer resolutions to be captured. Such more finely resolved studies provide
enough information to understand the late time behaviour, in particular, up to a 
time $t\simeq 4.2 \Lambda$.

\subsection{Event Horizon Finder}\label{evhorfinder}

\begin{figure}[t]
\begin{center}
\includegraphics[width=2.9in]{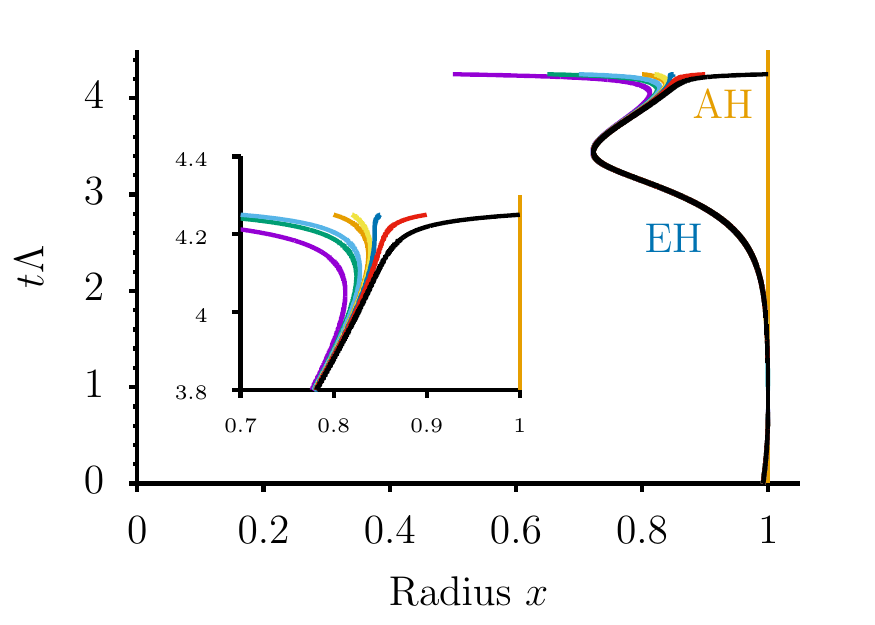}
\caption{Illustration of representative null rays traced to find
the event horizon. The inset shows a zoom-in at late times which aids
to visualize how null rays starting at different locations converge
as they are traced backwards in time.}
\label{fig:unstable_radius}
\end{center}
\end{figure}

To find the event horizon we trace null geodesics at late times back in time and determine
the surface $R(t)$ where they converge. To do so, we start from 
\begin{equation}
g_{ab}n^an^b=0 \;,
\label{eq:null_geo}
\end{equation}
where $n^a$ denotes the null tangent vector to the geodesics. 
Using \eqref{bgr} and the field redefinition
\eqref{redef} and \eqref{fieldsor} this relation implies 
\begin{equation}
\frac{\mathrm{d}x}{\mathrm{d}t} = -x^2 \left(a(t,x) + \frac{1}{2}\sigma(t,x)^2
\right) - x\sigma(t,x) - \frac{1}{2}\;,
\end{equation}
which we solve numerically using either a RK4 integrator or an second order
implicit integrator. The results obtained with both methods converge and are in excellent agreement. 
As described briefly above, we consider a collection of starting points at different radii and
bisect the resulting behavior to home-in on $R(t)$. Figure \ref{fig:unstable_radius} displays
eight representative initial conditions and illustrate the convergent behavior towards the event horizon.

\section{Appendix: Bounded scalar potentials in exotic holographic model}\label{secbounded}

In this section we explore modification of the model \eqref{s4} where the potential for the gravitational scalar 
$\chi$ is bounded; specifically we modify the nonlinear interactions between $\phi$ and $\chi$ as follows
\begin{equation}
-g\phi^2\chi^2\qquad \to \qquad -g\phi^2\chi^2 \left(1-f\ \chi^2 \right) \;,
\eqlabel{boundpoten}
\end{equation}
where $f={\rm const}>0$ is a new parameter. It is straightforward to modify the numerical code to reflect the change \eqref{boundpoten}.
We performed various tests and verified convergence of the new numerical code. In what follows we report the results of the analysis.

\begin{figure}[t]
\begin{center}
\includegraphics[width=2.9in]{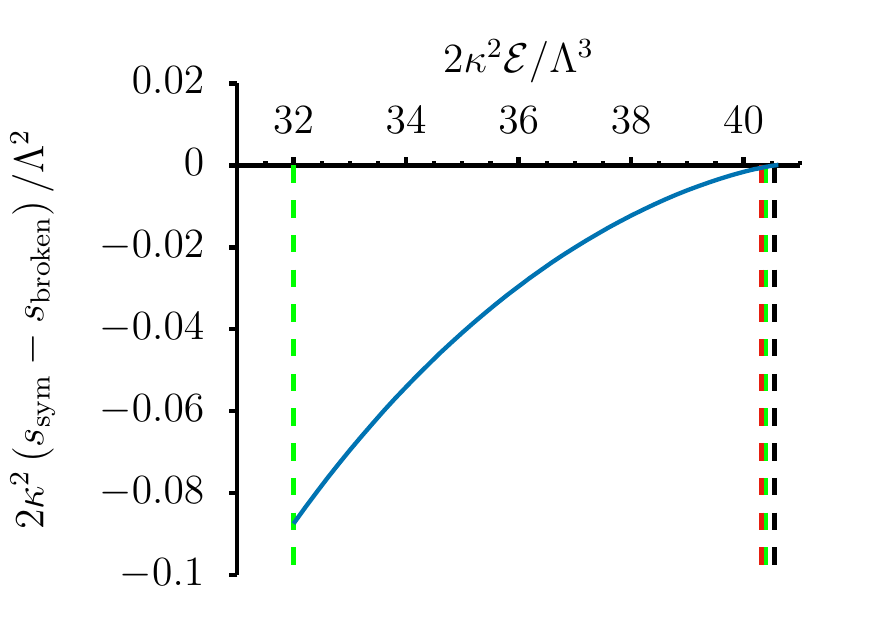}\hfill
\includegraphics[width=2.9in]{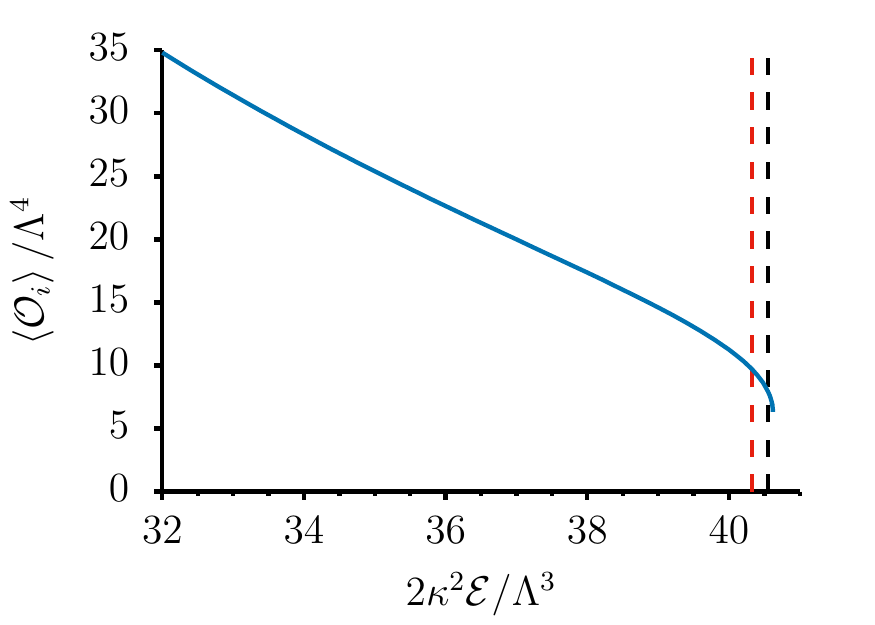}
\end{center}
 \caption{There is a new symmetry broken phase of the model \eqref{s4} for the modified 
nonlinear interaction \eqref{boundpoten}; here,
 $f=55$.  The left panel shows the difference in the entropy densities between symmetric and broken phases 
as a function of energy density. The (red) vertical dashed line is the onset of the linearized instability 
of the symmetric phase at $\cale=\cale_{crit}$, see \eqref{ecrit}. The (black) vertical dashed line 
denotes a new first-order phase transition at $\cale=\cale_{black}$, 
see \eqref{sblack}. The vertical (green) dashed lines 
indicate energy densities used in numerical evolutions, 
$\cale_{green,left}<\cale_{crit}<\cale_{green,right}<\cale_{black}$. The right panel represents the 
order parameter of the broken phase as a function of the energy density.} \label{appendixb1}
\end{figure}

Because modification \eqref{boundpoten} is a higher-order
$\chi$-nonlinear interaction, the linearized stability analyses are
not affected --- there is a linearized instability for
$\cale<\cale_{crit}$ with $\cale_{crit}$ given by \eqref{ecrit}.
Likewise, the static exotic branch bifurcating from the symmetric
phase at the onset of the instability is qualitatively unchanged, see
figure~\ref{figure2}.  However, for a wide range of $f>0$ we found a new
phase of the model with $\langle \calo_i\rangle \ne 0$. This new phase
exists for $\cale\lessgtr \cale_{crit}$, though it is numerically
challenging to find it as $f$ decreases and $\cale\to \cale_{crit}$.
The new branch enters a full phase diagram of the model in a fairly
complicated fashion.  For the results in figure~\ref{appendixb1} we
choose $f=55$. The left panel shows the entropy density difference
between the symmetric phase $s_{sym}$ and a new symmetry broken phase
$s_{broken}$.  The  (red) dashed vertical line identifies the onset of
the linearized instability at $\cale=\cale_{crit}$.  This new phase
dominates the microcanonical ensemble all the way to $\cale_{black}$,
denoted by the (black) vertical dashed line,
\begin{equation}
s_{broken}(\cale) > s_{sym}(\cale)\,,\qquad {\rm for}\qquad \cale< \cale_{black}=1.0057(3)\
 \cale_{crit} \;.
\eqlabel{sblack}
\end{equation}    
At $\cale=\cale_{black}$ there is a first-order phase transition, and since the symmetric phase at this energy density 
is perturbatively stable, the transition would occur dynamically only if the amplitude of the symmetry breaking fluctuations 
is large enough --- we explore this below for the energy density represented by the right (green) vertical dashed line
\begin{equation}
\cale_{crit}\qquad < \qquad \cale_{green,right}=1.0019(7)\ \cale_{crit} \qquad
<\qquad \cale_{black} \;.
\eqlabel{egreen}
\end{equation}  
The left (green) vertical dashed line corresponds to the energy
density \eqref{esymsim1}, $\cale_{green,left} < \cale_{crit}$.  The
right panel in figure~\ref{appendixb1} shows the order parameter,
$\langle \calo_i\rangle$, for the new symmetry breaking phase as a
function of the energy density $\cale$.

\begin{figure}[t]
\begin{center}
\includegraphics[width=2.9in]{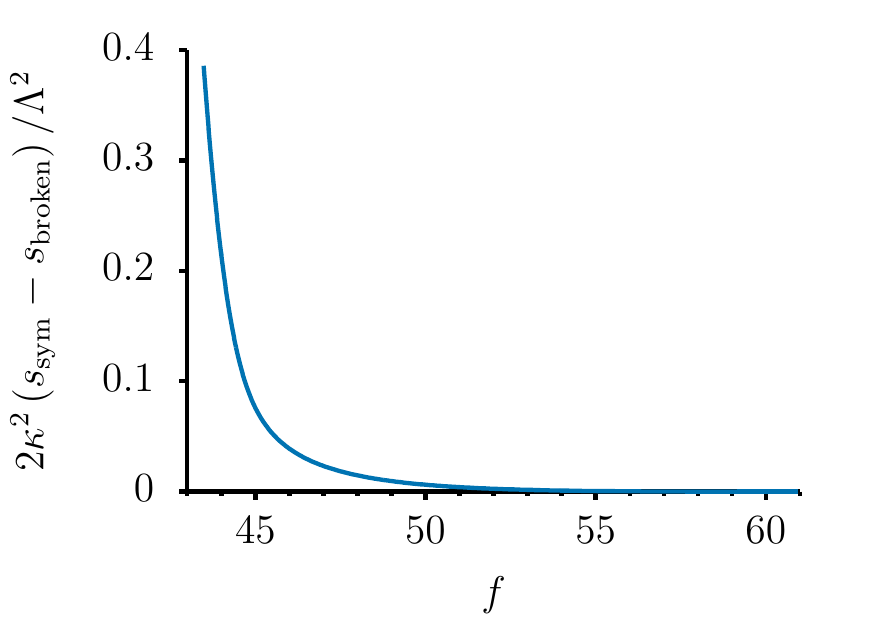}\hfill
\includegraphics[width=2.9in]{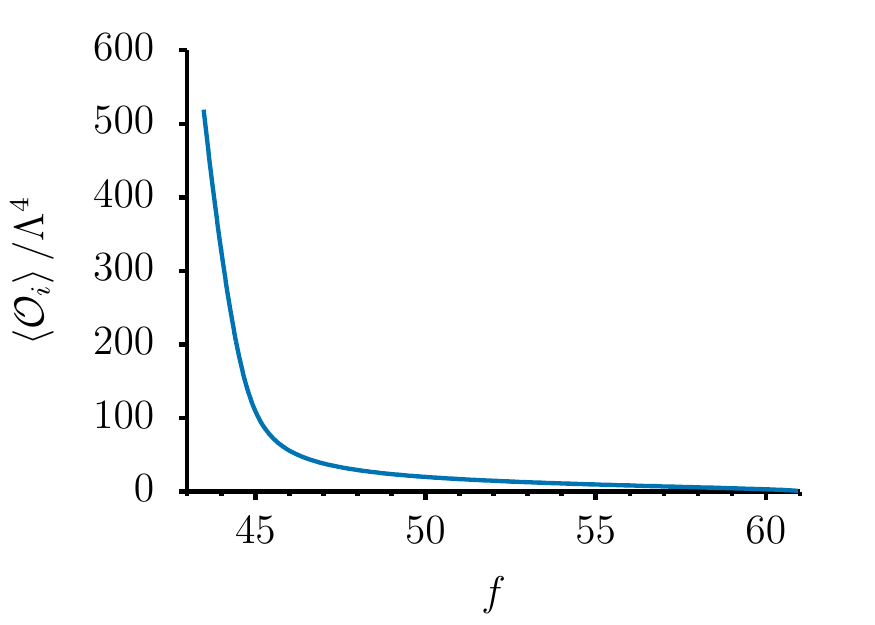}
\end{center}
 \caption{New symmetry broken phase dominates the microcanonical ensemble 
at  $\cale=\cale_{crit}$ for a wide range of the nonlinear parameter $f$,
``bounding'' the scalar potential in \eqref{boundpoten}. Right panel shows the corresponding dependence of the 
order parameter $\langle\calo_i\rangle$ in the broken phase. } \label{appendixb2}
\end{figure}

As figure~\ref{appendixb2} shows, the new symmetry broken phase dominates the 
microcanonical ensemble at $\cale=\cale_{crit}$ for all values of the $f$ in \eqref{boundpoten}
we studied. Notice that as $f$ decreases the new phase becomes very different from the symmetric phase:
it is much strongly favored entropically, and the symmetry breaking order parameter $\langle\calo_i\rangle$ (right panel) exhibits 
a rapid growth. All this is suggests that the limit $f\to 0_+$ is a singular one, as expected from the 
main text analysis of the $f=0$ model \eqref{s4}.

\begin{figure}[t]
\begin{center}
\includegraphics[width=2.9in]{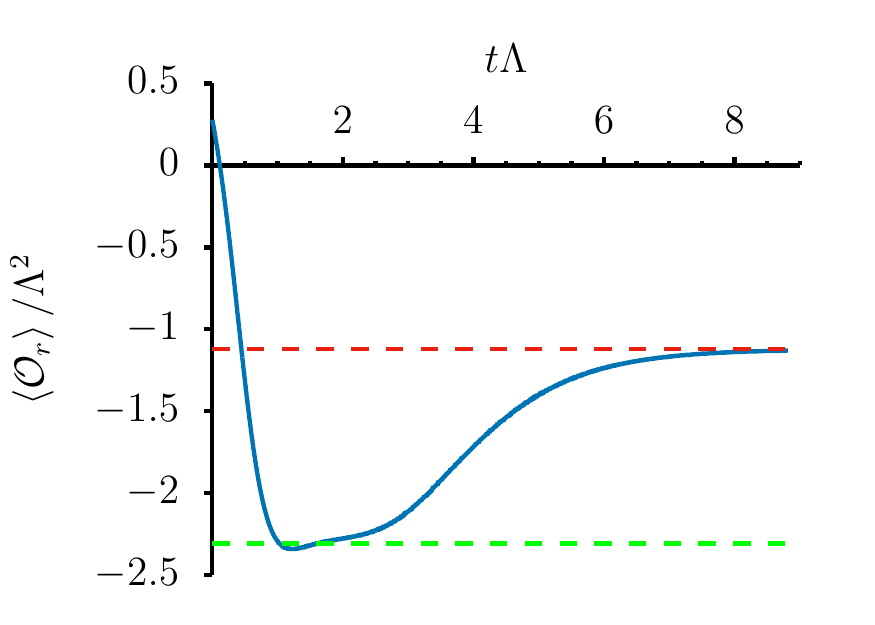} \hfill
\includegraphics[width=2.9in]{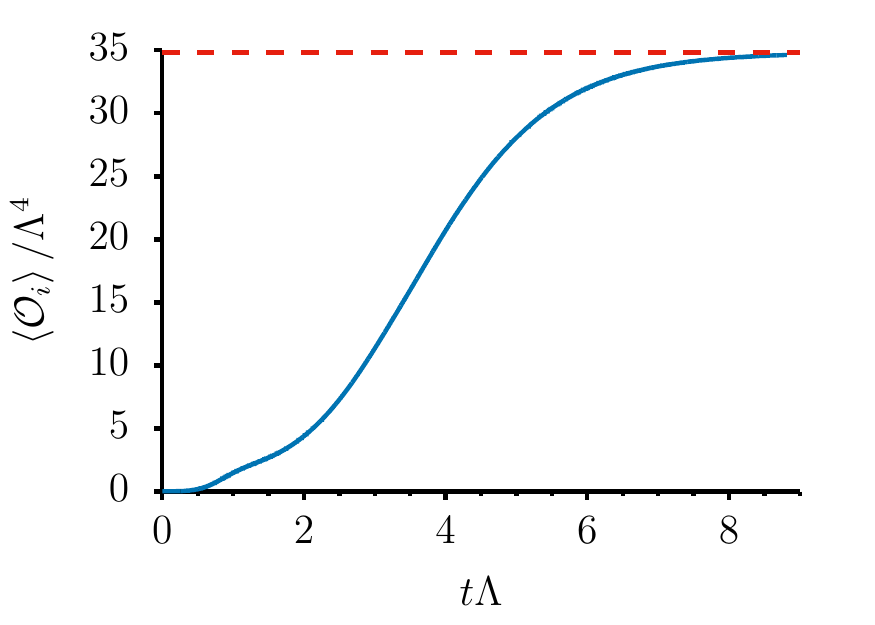}
\end{center}
 \caption{ Time evolution of the symmetric preserving order parameter 
$\langle\calo_r\rangle$ (left panel) and the symmetry breaking order parameter $\langle\calo_i\rangle$
(right panel) for $\cale<\cale_{crit}$, see \eqref{esymsim1}, 
and $f=55$. The system equilibrates to appropriate static values of the 
condensates, represented by (red) dashed  lines. The (green) dashed  line is the expectation 
value of $\calo_r$ in the symmetric phase at the corresponding energy density.} \label{appendixb3}
\end{figure}

Figure~\ref{appendixb3} represents the time evolution of model with $f=55$, $\cale=\cale_{green,left}=0.793642\ \cale_{crit}$, 
and the initial conditions 
chosen following Appendix \ref{initial} with values of $\cala_{p}$ and $\cala_{q}$ as in the simulations reported in 
section \ref{fullsim}. The symmetry preserving $\langle\calo_r\rangle$ (left panel) and the symmetry breaking 
$\langle\calo_i\rangle$ (right panel) condensates equilibrate to static values [(red) dashed lines] 
corresponding to the new symmetry broken phase discussed here. This should be contrasted with the $f=0$ results 
reported in section \ref{fullsim}, where the system evolves to a naked singularity.    
We initiate evolution with small amplitude of the symmetry breaking fluctuation
\begin{equation}
\langle\calo_i\rangle\bigg|_{t=0}=0.0252\ \Lambda^4 \;;
\eqlabel{oibelow}
\end{equation}
hence, they do not have enough time to become nonlinear at $t\Lambda \sim 1$, and the symmetric condensate 
$\langle\calo_r\rangle$ is close to its value in symmetry preserving phase at the corresponding energy density 
(represented by (green) dashed  line). For $t\Lambda >1 $ the symmetry breaking fluctuations continue to grow, ultimately 
capping off at the new equilibrium value.

\begin{figure}[t]
\begin{center}
\includegraphics[width=2.9in]{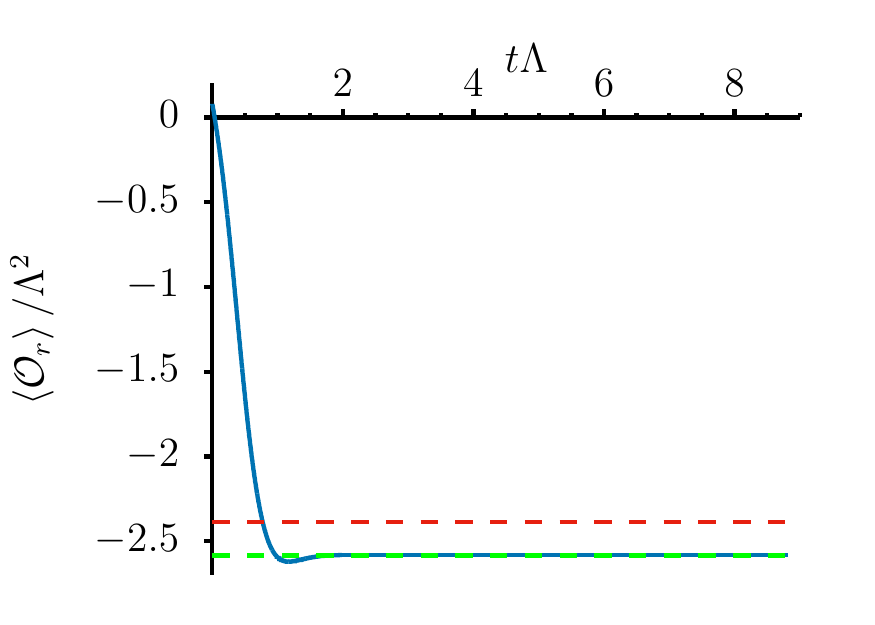}\hfill
\includegraphics[width=2.9in]{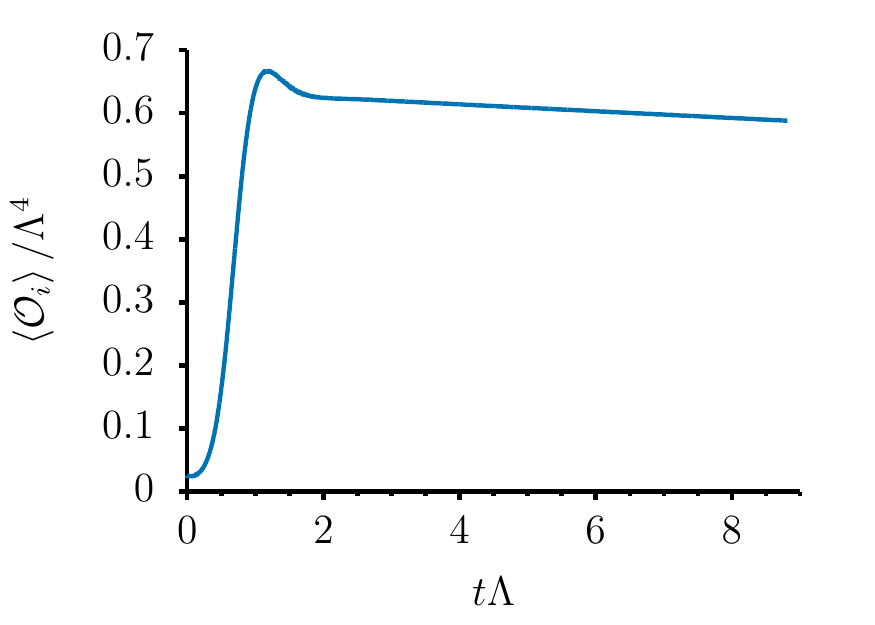}
\end{center}
 \caption{Dynamics of the model with $f=55$ and $\cale=\cale_{green,right}$, see \eqref{egreen} with 
initially small amplitude of the symmetry breaking fluctuations (right panel). Symmetry preserving 
condensate $\langle\calo_r\rangle$ equilibrates to the value in the symmetric phase [(green) dashed  line].
The (red) dashed  line is the value of this condensate at the same energy in the symmetry broken phase.} \label{appendixb4}
\end{figure}

\begin{figure}[t]
\begin{center}
\includegraphics[width=2.9in]{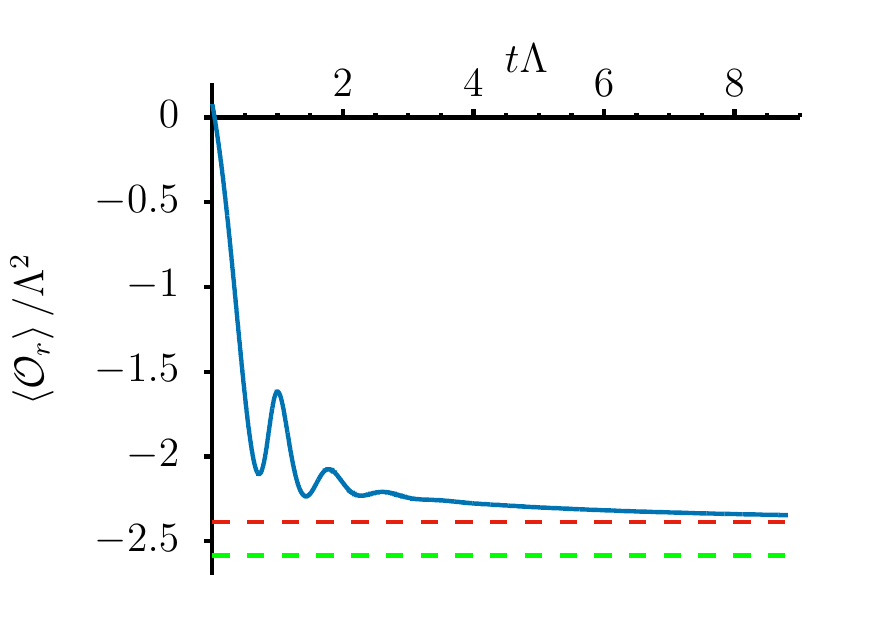} \hfill
\includegraphics[width=2.9in]{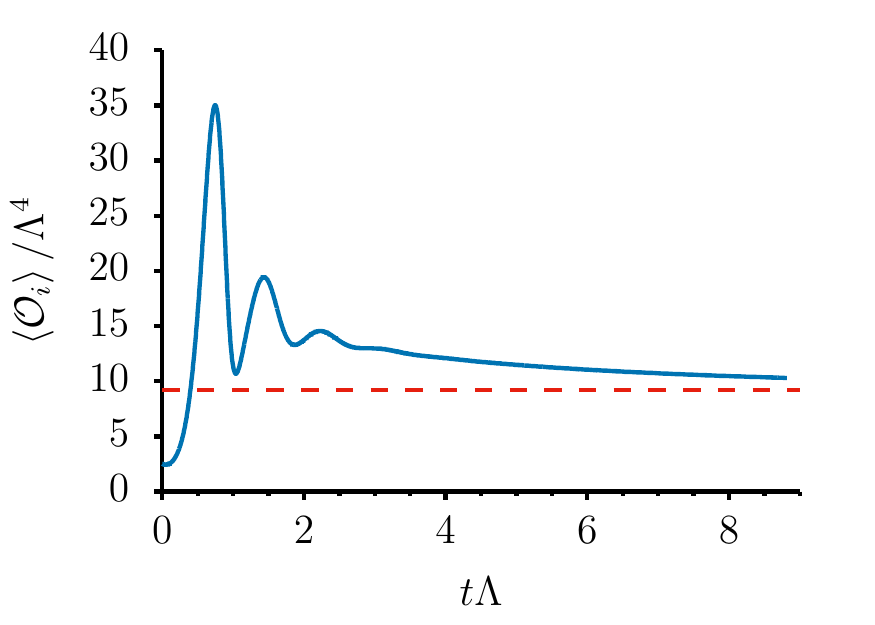}
\end{center}
 \caption{Dynamics of the model with $f=55$ and $\cale=\cale_{green,right}$, see \eqref{egreen} with 
initially large amplitude of the symmetry breaking fluctuations (right panel). Symmetry preserving 
condensate $\langle\calo_r\rangle$ approaches the equilibrium value in the symmetry broken 
phase [ (red) dashed line].
The (green) dashed  line is the value of this condensate at the same energy in the symmetric phase.} \label{appendixb5}
\end{figure}

As shown in figure~\ref{appendixb1}, the new symmetry broken phase has an interesting feature in the narrow 
energy range:
\begin{equation}
\cale_{crit}\qquad <\qquad  \cale\qquad <\qquad \cale_{black} \;.
\eqlabel{enrange}
\end{equation}
Here, the symmetric phase is perturbatively stable, but the new symmetry broken phase is nonetheless 
entropically favorable; thus one excepts that the broken phase can be reached dynamically only if the 
amplitude of the initial symmetry breaking fluctuations is sufficiently large. 
We find that this is indeed the case. For the results presented in figure~\ref{appendixb4}, 
$\langle\calo_i\rangle\bigg|_{t=0}=0.0252\ \Lambda^4$ and the system equilibrates to a (metastable) 
symmetric phase. Figure~\ref{appendixb5} represents results of the simulation 
for the initial condition with $\langle\calo_i\rangle\bigg|_{t=0}=2.52\ \Lambda^4$ --- here the amplitude 
is large enough to reach the entropically dominant symmetry broken phase. The approach to equilibrium in both 
cases is very slow as the energy density of the simulations is close to the critical one, see \eqref{egreen}.

\bibliographystyle{h-physrev.bst}
\bibliography{unstable.bib}

\begin{thebibliography}{10}

\bibitem{m1}
J.~M. Maldacena,
\newblock Int. J. Theor. Phys. {\bf 38}, 1113 (1999), hep-th/9711200,
\newblock [Adv. Theor. Math. Phys.2,231(1998)].

\bibitem{Aharony:1999ti}
O.~Aharony, S.~S. Gubser, J.~M. Maldacena, H.~Ooguri, and Y.~Oz,
\newblock Phys. Rept. {\bf 323}, 183 (2000), hep-th/9905111.

\bibitem{Hartnoll:2008vx}
S.~A. Hartnoll, C.~P. Herzog, and G.~T. Horowitz,
\newblock Phys. Rev. Lett. {\bf 101}, 031601 (2008), 0803.3295.

\bibitem{Hartnoll:2008kx}
S.~A. Hartnoll, C.~P. Herzog, and G.~T. Horowitz,
\newblock JHEP {\bf 0812}, 015 (2008), 0810.1563.

\bibitem{Klebanov:1999tb}
I.~R. Klebanov and E.~Witten,
\newblock Nucl. Phys. B {\bf 556}, 89 (1999), hep-th/9905104.

\bibitem{Gregory:1993vy}
R.~Gregory and R.~Laflamme,
\newblock Phys. Rev. Lett. {\bf 70}, 2837 (1993), hep-th/9301052.

\bibitem{Hartnoll:2009sz}
S.~A. Hartnoll,
\newblock Class. Quant. Grav. {\bf 26}, 224002 (2009), 0903.3246.

\bibitem{Buchel:2009ge}
A.~Buchel and C.~Pagnutti,
\newblock Nucl. Phys. B {\bf 824}, 85 (2010), 0904.1716.

\bibitem{Donos:2011ut}
A.~Donos and J.~P. Gauntlett,
\newblock JHEP {\bf 1106}, 053 (2011), 1104.4478.

\bibitem{Hubeny:2002xn}
V.~E. Hubeny and M.~Rangamani,
\newblock JHEP {\bf 0205}, 027 (2002), hep-th/0202189.

\bibitem{Buchel:2015gxa}
A.~Buchel and L.~Lehner,
\newblock Class. Quant. Grav. {\bf 32}, 145003 (2015), 1502.01574.

\bibitem{Buchel:2015pla}
A.~Buchel,
\newblock 1509.07780.

\bibitem{Dias:2015pda}
O.~J.~C. Dias, J.~E. Santos, and B.~Way,
\newblock JHEP {\bf 1504}, 060 (2015), 1501.06574.

\bibitem{Chesler:2013lia}
P.~M. Chesler and L.~G. Yaffe,
\newblock JHEP {\bf 1407}, 086 (2014), 1309.1439.

\bibitem{Gubser:2005ih}
S.~S. Gubser,
\newblock Class. Quant. Grav. {\bf 22}, 5121 (2005), hep-th/0505189.

\bibitem{Buchel:2010wk}
A.~Buchel and C.~Pagnutti,
\newblock Phys. Lett. B {\bf 697}, 168 (2011), 1010.5748.

\bibitem{Gubser:2000ec}
S.~S. Gubser and I.~Mitra,
\newblock hep-th/0009126.

\bibitem{Gubser:2000mm}
S.~S. Gubser and I.~Mitra,
\newblock JHEP {\bf 0108}, 018 (2001), hep-th/0011127.

\bibitem{1962RSPSA.269...21B}
H.~{Bondi}, M.~G.~J. {van der Burg}, and A.~W.~K. {Metzner},
\newblock {Proceedings of the Royal Society of London Series A} {\bf 269}, 21
  (1962).

\bibitem{Bishop:1997ik}
N.~T. Bishop, R.~Gomez, L.~Lehner, M.~Maharaj, and J.~Winicour,
\newblock Phys. Rev. D {\bf 56}, 6298 (1997), gr-qc/9708065.

\bibitem{2012LRR....15....2W}
J.~{Winicour},
\newblock Living Reviews in Relativity {\bf 15}, 2 (2012), 0810.1903.

\bibitem{Booth:2005qc}
I.~Booth,
\newblock Can. J. Phys. {\bf 83}, 1073 (2005), gr-qc/0508107.

\bibitem{Figueras:2009iu}
P.~Figueras, V.~E. Hubeny, M.~Rangamani, and S.~F. Ross,
\newblock JHEP {\bf 0904}, 137 (2009), 0902.4696.

\bibitem{Buchel:2010wp}
A.~Buchel,
\newblock Nucl. Phys. B {\bf 847}, 297 (2011), 1012.2404.

\bibitem{Bosch:2016vcp}
P.~Bosch, S.~R. Green, and L.~Lehner,
\newblock Phys. Rev. Lett. {\bf 116}, 141102 (2016), 1601.01384.

\bibitem{Crisford:2017zpi}
T.~Crisford and J.~E. Santos,
\newblock 1702.05490.

\bibitem{bcy}
A.~Buchel, P.~M. Chesler, and L.~G. Yaffe,
\newblock in progress.

\bibitem{ms}
M.~Srednicki,
\newblock Phys. Rev. E {\bf 50}, 888 (1994), cond-mat/9403051.

\bibitem{Bizon:2011gg}
P.~Bizon and A.~Rostworowski,
\newblock Phys. Rev. Lett. {\bf 107}, 031102 (2011), 1104.3702.

\bibitem{Buchel:2012uh}
A.~Buchel, L.~Lehner, and S.~L. Liebling,
\newblock Phys. Rev. D {\bf 86}, 123011 (2012), 1210.0890.

\bibitem{Buchel:2013uba}
A.~Buchel, S.~L. Liebling, and L.~Lehner,
\newblock Phys. Rev. D {\bf 87}, 123006 (2013), 1304.4166.

\bibitem{Balasubramanian:2014cja}
V.~Balasubramanian, A.~Buchel, S.~R. Green, L.~Lehner, and S.~L. Liebling,
\newblock Phys. Rev. Lett. {\bf 113}, 071601 (2014), 1403.6471.

\bibitem{Green:2015dsa}
S.~R. Green, A.~Maillard, L.~Lehner, and S.~L. Liebling,
\newblock Phys. Rev. D {\bf 92}, 084001 (2015), 1507.08261.

\bibitem{Dias:2016eto}
O.~J.~C. Dias, J.~E. Santos, and B.~Way,
\newblock Phys. Rev. Lett. {\bf 117}, 151101 (2016), 1605.04911.

\bibitem{Withers:2013loa}
B.~Withers,
\newblock Class. Quant. Grav. {\bf 30}, 155025 (2013), 1304.0129.

\bibitem{Pilch:2000ue}
K.~Pilch and N.~P. Warner,
\newblock Nucl. Phys. B {\bf 594}, 209 (2001), hep-th/0004063.

\bibitem{Buchel:2000cn}
A.~Buchel, A.~W. Peet, and J.~Polchinski,
\newblock Phys. Rev. D {\bf 63}, 044009 (2001), hep-th/0008076.

\bibitem{Buchel:2013id}
A.~Buchel, J.~G. Russo, and K.~Zarembo,
\newblock JHEP {\bf 1303}, 062 (2013), 1301.1597.

\bibitem{Garfinkle:2003bb}
D.~Garfinkle,
\newblock Phys. Rev. Lett. {\bf 93}, 161101 (2004), gr-qc/0312117.

\bibitem{Buchel:2014gta}
A.~Buchel, R.~C. Myers, and A.~van Niekerk,
\newblock JHEP {\bf 1502}, 017 (2015), 1410.6201,
\newblock Erratum: [JHEP 1507, 137 (2015)], 10.1007/JHEP07(2015)137.

\end{thebibliography}

\end{document}